\input harvmac
\noblackbox
\input epsf
\def\abstract#1{
\vskip .5in\vfil\centerline
{\bf Abstract}\penalty1000
{{\smallskip\ifx\answ\bigans\leftskip 1pc \rightskip 1pc 
\else\leftskip 1pc \rightskip 1pc\fi
\noindent \abstractfont  \baselineskip=12pt
{#1} \smallskip}}
\penalty-1000}
%
\def\hth/#1#2#3#4#5#6#7{{\tt hep-th/#1#2#3#4#5#6#7}}
\def\nup#1({Nucl.\ Phys.\ $\us {B#1}$\ (}
\def\plt#1({Phys.\ Lett.\ $\us  {B#1}$\ (}
\def\cmp#1({Comm.\ Math.\ Phys.\ $\us  {#1}$\ (}
\def\prp#1({Phys.\ Rep.\ $\us  {#1}$\ (}
\def\prl#1({Phys.\ Rev.\ Lett.\ $\us  {#1}$\ (}
\def\prv#1({Phys.\ Rev.\ $\us  {#1}$\ (}
\def\mpl#1({Mod.\ Phys.\ Let.\ $\us  {A#1}$\ (}
\def\atmp#1({Adv.\ Theor.\ Math.\ Phys.\ $\us  {#1}$\ (}
\def\ijmp#1({Int.\ J.\ Mod.\ Phys.\ $\us{A#1}$\ (}
\def\jhep#1({JHEP\ $\us {#1}$\ (}

\def\subsubsec#1{\vskip 0.2cm \goodbreak \noindent {\it #1} \br}

\def\bb#1{{\bar{#1}}}
\def\bx#1{{\bf #1}}
\def\cx#1{{\cal #1}}
\def\tx#1{{\tilde{#1}}}
\def\hx#1{{\hat{#1}}}
\def\rmx#1{{\rm #1}}
\def\us#1{\underline{#1}}
\def\fc#1#2{{#1\over #2}}
\def\frac#1#2{{#1\over #2}}

\def\br{\hfill\break}
\def\ni{\noindent}

\def\al{\alpha}\def\be{\beta}\def\ga{\gamma}\def\om{\omega}
\def\p{\partial}

\def\ZZ{{\bx Z}}
\def\la{\langle}\def\ra{\rangle}

\def\IP{{\bx P}}\def\WP{{\bx {WP}}}
\def\ss{\scriptstyle}
\def\lrai#1{\lra \hskip-0.5cm ^{#1} \hskip0.4cm}
\def\OO#1{\cx O{\ss #1}}
\def\fff{~}
\def\htopop{\cx H^{top}_{op}}\def\htopcl{\cx H^{top}_{cl}}
\def\htopopz{\cx H^{top,0}_{op}}

\def\lra{\, \longrightarrow\, }
\def\pmut{\underline{P}}\def\lmut{\underline{L}}\def\rmut{\underline{R}}

\def\CC{{\bx C}}\def\CY{Calabi--Yau\ }\def\LG{Landau--Ginzburg\ }
\def\ch{{\rm ch}}\def\td{{\rm td}}\def\Ext{{\rm Ext}}
\def\derc{D^\flat}
\def\mchi{\chi}
\def\tX{{\tilde{X}}}
\def\bR{\{R_a\}}\def\bS{\{S^a\}}\def\bSd{\{S^{a\, *}\}}

\def\hoo{{h_{1,1}}}\def\Piv{\vec{\Pi}}\def\Qv{\vec{Q}}
\def\jb{\bar{j}}\def\psib{\psi}
\lref\rTis{
S.~Hosono, A.~Klemm, S.~Theisen and S.~T.~Yau,
``Mirror symmetry, mirror map and applications to complete intersection  Calabi-Yau spaces,''
Nucl.\ Phys.\  {\bf B433}, 501 (1995)
hep-th/9406055.
}

\lref\rHaM{
J.~A.~Harvey and G.~Moore,
``On the algebras of BPS states,''
Commun.\ Math.\ Phys.\  {\bf 197}, 489 (1998)
hep-th/9609017.
}
\lref\rBat{V. Batyrev, 
``Dual Polyhedra and Mirror Symmetry for Calabi-Yau Hypersurfaces in Toric Varieties,''
Journ. Alg. Geom. 3 (1994) 493; Duke Math. Journ.
69 (1993) 349.}

\lref\rWKth{
E.~Witten,
``D-branes and K-theory,''
JHEP {\bf 9812}, 019 (1998)
hep-th/9810188.
}

\lref\rcandii{
P.~Candelas, A.~Font, S.~Katz and D.~R.~Morrison,
``Mirror symmetry for two parameter models. 2,''
Nucl.\ Phys.\  {\bf B429}, 626 (1994)
hep-th/9403187.
}
\lref\rigsm{
S.~Govindarajan and T.~Jayaraman,
``D-branes, Exceptional Sheaves and Quivers on Calabi-Yau manifolds: From Mukai to McKay,''
hep-th/0010196;\ 
A. Tomasiello, ``D-branes on Calabi--Yau manifolds and helices,''
hep-th/0010217.
}

\lref\rCYDbsTG{
C.~Beasley, B.~R.~Greene, C.~I.~Lazaroiu and M.~R.~Plesser,
``D3-branes on partial resolutions of abelian quotient singularities of  Calabi-Yau threefolds,''
Nucl.\ Phys.\  {\bf B566}, 599 (2000)
hep-th/9907186;\
B.~Feng, A.~Hanany and Y.~He,
``D-brane gauge theories from toric singularities and toric duality,''
hep-th/0003085;\ 
P.~S.~Aspinwall and M.~R.~Plesser,
``D-branes, discrete torsion and the McKay correspondence,''
hep-th/0009042;\
}

\lref\rms{``Essays on Mirror Manifolds'', (ed. S.T. Yau), 
Int. Press, Hong Kong, 1992.}

\lref\rGJS{
S.~Govindarajan, T.~Jayaraman and T.~Sarkar,
``Worldsheet approaches to D-branes on supersymmetric cycles,''
Nucl.\ Phys.\  {\bf B580}, 519 (2000)
hep-th/9907131.
}

\lref\CAFP{
A.~Ceresole, R.~D'Auria, S.~Ferrara and A.~Van Proeyen,
``Duality transformations in supersymmetric Yang-Mills theories coupled to supergravity,''
Nucl.\ Phys.\  {\bf B444}, 92 (1995)
hep-th/9502072.
}
\lref\rMms{
D.~R.~Morrison,
``The Geometry Underlying Mirror Symmetry,''
alg-geom/9608006.
}
\lref\rDtor{
M.~R.~Douglas,
``D-branes and discrete torsion,''
hep-th/9807235;\
M.~R.~Douglas and B.~Fiol,
``D-branes and discrete torsion. II,''
hep-th/9903031.
}

\lref\rCVtat{
S.~Cecotti and C.~Vafa,
Nucl.\ Phys.\  {\bf B367}, 359 (1991).
}

\lref\rCVclass{
S.~Cecotti and C.~Vafa,
Commun.\ Math.\ Phys.\  {\bf 158}, 569 (1993)
hep-th/9211097.
}

\lref\rHos{
S.~Hosono,
``Local mirror symmetry and type IIA monodromy of Calabi-Yau manifolds,''
hep-th/0007071.
}

\lref\rBD{
V.~V.~Batyrev and D.~I.~Dais,
``Strong McKay correspondence, string theoretic Hodge numbers and mirror symmetry,''
alg-geom/9410001.
}

\lref\rDG{
D.~Diaconescu and J.~Gomis,
``Fractional branes and boundary states in orbifold theories,''
JHEP {\bf 0010}, 001 (2000)
hep-th/9906242.
}
\lref\rCYDbii{
P.~Kaste, W.~Lerche, C.~A.~Lutken and J.~Walcher,
``D-branes on K3-fibrations,''
Nucl.\ Phys.\  {\bf B582}, 203 (2000)
 hep-th/9912147;\ 
J.~Fuchs, P.~Kaste, W.~Lerche, C.~A.~Lutken, C.~Schweigert and J.~Walcher,
``Boundary fixed points, enhanced gauge symmetry and singular bundles  on K3,''
hep-th/0007145;\ 
W.~Lerche, C.~A.~Lutken and C.~Schweigert,
``D-branes on ALE spaces and the ADE classification of conformal field  theories,''
hep-th/0006247;\ 
W.~Lerche,
``On a boundary CFT description of nonperturbative N = 2 Yang-Mills  theory,''
hep-th/0006100.}

\lref\rSch{
E.~Scheidegger,
``D-branes on some one- and two-parameter Calabi-Yau hypersurfaces,''
JHEP {\bf 0004}, 003 (2000)
 hep-th/9912188;\ 
}

\lref\rDFR{
M.~R.~Douglas, B.~Fiol and C.~R\"omelsberger,
``The spectrum of BPS branes on a noncompact Calabi-Yau,''
hep-th/0003263;\ 
``Stability and BPS branes,''
hep-th/0002037.
}

\lref\rHV{
K.~Hori and C.~Vafa,
``Mirror symmetry,''
hep-th/0002222.}

\lref\rHIV{
K.~Hori, A.~Iqbal and C.~Vafa,
``D-branes and mirror symmetry,''
hep-th/0005247.
}

\lref\rOOY{
H.~Ooguri, Y.~Oz and Z.~Yin,
``D-branes on Calabi-Yau spaces and their mirrors,''
Nucl.\ Phys.\  {\bf B477}, 407 (1996)
 hep-th/9606112.
}

\lref\rCYDbscft{
M.~Gutperle and Y.~Satoh,
``D-branes in Gepner models and supersymmetry,''
Nucl.\ Phys.\  {\bf B543}, 73 (1999)
 hep-th/9808080;\ 
S.~Govindarajan and T.~Jayaraman,
``On the Landau-Ginzburg description of boundary CFTs and special  Lagrangian submanifolds,''
JHEP {\bf 0007}, 016 (2000)
hep-th/0003242;\ 
I.~Brunner and V.~Schomerus,
``On superpotentials for D-branes in Gepner models,''
hep-th/0008194.
}

\lref\rGJlsm{
S.~Govindarajan, T.~Jayaraman and T.~Sarkar,
``On D-branes from gauged linear sigma models,''
hep-th/0007075;\ 
}

\lref\rRS{
A.~Recknagel and V.~Schomerus,
``D-branes in Gepner models,''
Nucl.\ Phys.\  {\bf B531}, 185 (1998)
 hep-th/9712186.
}

\lref\rDD{
D.~Diaconescu and M.~R.~Douglas,
``D-branes on stringy Calabi-Yau manifolds,''
hep-th/0006224.
}

\lref\rDR{
D.~Diaconescu and C.~R\"omelsberger,
``D-branes and bundles on elliptic fibrations,''
Nucl.\ Phys.\  {\bf B574}, 245 (2000)
 hep-th/9910172.
}

\lref\rMDrev{
M.~R.~Douglas,
``D-branes on Calabi-Yau manifolds,''
 math.ag/0009209.
}

\lref\rDQ{
I.~Brunner, M.~R.~Douglas, A.~Lawrence and C.~R\"omelsberger,
``D-branes on the quintic,''
JHEP {\bf 0008}, 015 (2000)
 hep-th/9906200.
}

\lref\rNW{
N.~P.~Warner,
``Supersymmetry in boundary integrable models,''
Nucl.\ Phys.\  {\bf B450}, 663 (1995)
 hep-th/9506064.
}

\lref\rcand{
P.~Candelas, X.~C.~De La Ossa, P.~S.~Green and L.~Parkes,
``A pair of Calabi-Yau manifolds as an exactly soluble superconformal  theory,''
Nucl.\ Phys.\  {\bf B359}, 21 (1991).
}

\lref\rSemip{
A.~C.~Avram, E.~Derrick and D.~Jancic,
``On Semi-Periods,''
Nucl.\ Phys.\  {\bf B471}, 293 (1996)
 hep-th/9511152.
}

\lref\rWlsm{
E.~Witten,
``Phases of N = 2 theories in two dimensions,''
Nucl.\ Phys.\  {\bf B403}, 159 (1993)
 hep-th/9301042.
}

\lref\rSW{
N.~Seiberg and E.~Witten,
``Electric - magnetic duality, monopole condensation, and confinement in N=2 supersymmetric Yang-Mills theory,''
Nucl.\ Phys.\  {\bf B426}, 19 (1994)
 hep-th/9407087.
}

\lref\rLVW{
W.~Lerche, C.~Vafa and N.~P.~Warner,
``Chiral Rings In N=2 Superconformal Theories,''
Nucl.\ Phys.\  {\bf B324}, 427 (1989).
}

\lref\rCVLG{
C.~Vafa,
``String Vacua And Orbifoldized L-G Models,''
Mod.\ Phys.\ Lett.\  {\bf A4}, 1169 (1989);
``Topological Landau-Ginzburg models,''
Mod.\ Phys.\ Lett.\  {\bf A6}, 337 (1991).
}

\lref\rZas{
E.~Zaslow,
``Solitons and helices: The Search for a math physics bridge,''
Commun.\ Math.\ Phys.\  {\bf 175}, 337 (1996)
 hep-th/9408133.
}

\lref\rMPlsm{
For a formulation in terms of toric geometry, see:\br
D.~R.~Morrison and M.~Ronen Plesser,
``Summing the instantons: Quantum cohomology and mirror symmetry in toric varieties,''
Nucl.\ Phys.\  {\bf B440}, 279 (1995)
 hep-th/9412236,
}

\lref\rDM{
M.~R.~Douglas and G.~Moore,
``D-branes, Quivers, and ALE Instantons,''
hep-th/9603167.
}

\lref\rSD{
C.~M.~Hull and P.~K.~Townsend,
``Unity of superstring dualities,''
Nucl.\ Phys.\  {\bf B438}, 109 (1995)
 hep-th/9410167;\ 
E.~Witten,
``String theory dynamics in various dimensions,''
Nucl.\ Phys.\  {\bf B443}, 85 (1995)
 hep-th/9503124;\
A.~Strominger,
``Massless black holes and conifolds in string theory,''
Nucl.\ Phys.\  {\bf B451}, 96 (1995)
 hep-th/9504090;\ 
J.~Polchinski,
``Dirichlet-Branes and Ramond-Ramond Charges,''
Phys.\ Rev.\ Lett.\  {\bf 75}, 4724 (1995)
 hep-th/9510017.
}

\lref\rWindex{
E.~Witten,
``Constraints On Supersymmetry Breaking,''
Nucl.\ Phys.\  {\bf B202}, 253 (1982).
}

\lref\rRW{
R.~Rohm and E.~Witten,
``The Antisymmetric Tensor Field In Superstring Theory,''
Annals Phys.\  {\bf 170}, 454 (1986).
}

\lref\rDBcoups{
M.~B.~Green, J.~A.~Harvey and G.~Moore,
``I-brane inflow and anomalous couplings on D-branes,''
Class.\ Quant.\ Grav.\  {\bf 14}, 47 (1997)
 hep-th/9605033;\
R.~Minasian and G.~Moore,
``K-theory and Ramond-Ramond charge,''
JHEP {\bf 9711}, 002 (1997) 
 hep-th/9710230;\
Y.~E.~Cheung and Z.~Yin,
``Anomalies, branes, and currents,''
Nucl.\ Phys.\  {\bf B517}, 69 (1998)
 hep-th/9710206.
}

\lref\rFW{
D.~S.~Freed and E.~Witten,
``Anomalies in string theory with D-branes,''
hep-th/9907189.
}

\lref\rLW{
W.~Lerche and N.~P.~Warner,
``Index Theorems In N=2 Superconformal Theories,''
Phys.\ Lett.\  {\bf B205}, 471 (1988).
}
\lref\rVmc{
C.~Vafa,
``Extending mirror conjecture to Calabi-Yau with bundles,''
hep-th/9804131.
}

\lref\rDGM{
M.~R.~Douglas, B.~R.~Greene and D.~R.~Morrison,
``Orbifold resolution by D-branes,''
Nucl.\ Phys.\  {\bf B506}, 84 (1997)
 hep-th/9704151.
}

\lref\rIN{Y.~Ito and H.~Nakajima, ``McKay correspondence 
and Hilbert schemes in dimension three'', math.AG/9803120.}
\lref\rKon{M.~Kontsevich, ``Homological Algebra of Mirror Symmetry,''
alg-geom/9411018.}
\lref\rRudi{A.~N.~Rudakov (ed.), ``Helices and vector bundles:
Seminaire Rudakov'', London Mathematical Society Lecture Note Series 148,
Cambridge University Press, Cambridge 1990.}
\lref\rBon{A.~I.~Bondal, ``Helices, representations of quivers and
Koszul algebras'', in \rRudi; Math. USSR Izv. {\bf 34} (1990) 23.}


\vskip-2cm
\Title{\vbox{
\rightline{\vbox{\baselineskip12pt\hbox{CERN-TH/2000-315}
                            \hbox{hep-th/0010223}}}}}
{Phases of Supersymmetric D-branes on K\"ahler Manifolds}
\vskip -0.8cm

\centerline{\titlefont and the McKay correspondence}

\abstractfont 
%
\vskip 0.8cm
\centerline{P. Mayr}
\vskip 0.8cm
\centerline{CERN Theory Division} 
\centerline{CH-1211 Geneva 23}
\centerline{Switzerland}
\vskip 0.3cm
\abstract{%
We study the topological zero mode sector of type II strings
on a K\"ahler manifold $X$ in the presence of boundaries.
We construct two finite bases, in a sense bosonic and fermionic,
that generate the topological sector of the Hilbert space 
with boundaries. The fermionic basis localizes on compact submanifolds in $X$.
A variation of the FI terms interpolates between the description 
of these ground states in terms of the ring of chiral fields 
at the boundary at small volume and helices 
of exceptional sheaves at large volume, respectively. The 
identification of the bosonic/fermionic basis with the dual bases
for the non-compact/compact K-theory group on $X$ gives a 
natural explanation of the McKay correspondence in terms of a
linear sigma model and suggests a simple generalization 
of McKay to singular resolutions. The construction 
provides also a very effective way to describe D-brane states 
on generic, compact Calabi--Yau manifolds and allows to
recover detailed information on the moduli space, such as
monodromies and analytic continuation matrices, from the
group theoretical data of a simple orbifold.
}
\Date{\vbox{\hbox{ {October 2000}}
}}
\goodbreak

\parskip=4pt plus 15pt minus 1pt
\baselineskip=14pt 
\leftskip=8pt \rightskip=10pt
%

\newsec{Introduction}
The fact that open strings constitute an important sector 
of type II closed string theories has not been appreciated appropriately
before the seminal works on string duality \rSD, although it
appeared earlier, somehow in disguise, 
in the mathematical literature on mirror symmetry \rKon.
As the Dirichlet boundary conditions of the open strings break
half of the supersymmetry of the closed string sector, this
extends the beautiful geometric structure of the vacua 
of non-perturbative $\cx N=2$ supersymmetric theories 
\rSW\ described by closed strings, to $\cx N=1$ supersymmetry theories 
in terms of open strings.

It is therefore clearly very important to study 
this sector of type II strings which
keeps new aspects of mirror symmetry \rKon\rVmc\rMms\
and non-perturbative $\cx N=1$ physics. The two-dimensional 
perspective has been emphasized in the works \rOOY\rRS\rDQ\ and there has
been much conceptional progress since then \rDM\rDGM\rDtor\rDFR\rHIV\rDD,
see \rMDrev\ for a review and a more complete list of references.
In a first step one would like
to understand the zero mode structure, that is the 
open string ground states which represent BPS D-branes. Here we will
show that Witten's gauged linear sigma model \rWlsm\rMPlsm\ 
provides the natural language to construct a finite basis,
in a sense to be made precise, for 
all D-branes on a K\"ahler manifold $X$. We follow closely
the discussion of \rWlsm\ which in particular allows to 
interpolate between D-branes at large and small volume. In the
closed string case, the topological Hilbert space $\htopcl$ has a large
volume interpretation in terms of a deformation of the
cohomology ring of $X$ \rWindex\rRW\rLVW, and a small volume interpretation in 
terms of the ring of chiral fields of a $(2,2)$ LG  theory\foot{To 
simplify
notation we will loosely
refer to the small volume orbifold 
phase also as the \LG phase, irrespectively of the choice of the
superpotential.}.
We find a very similar structure in the topological sector
with boundaries, with the large volume phase described by
helices of 
exceptional sheaves, and the small volume phase by
the ring of zero modes of the chiral fields at the boundary. 

A key point is
that at the boundary, the chiral fields split into representations
of the unbroken supersymmetry with bosonic and fermionic statistics
in the directions transverse and normal to the brane,
respectively\foot{In fact many properties of the sector with
boundary are similar to the bulk sector of the theory with  $(0,2)$ 
world-sheet supersymmetry.}. 
Thus the ordinary ring of zero modes of 
chiral super-fields splits into a ring of bosons and a ring of fermions.
Multiplication of the trivial ground state 
with the bosonic zero modes leads to
a basis $\bR$ of generators for
the topological sector with boundaries $\htopop$ that correspond
to an exceptional collection of line bundles which spread over the total space $X$. 
Multiplication with fermionic zero modes generates
a dual basis $\bS$ of generators for $\htopop$ 
that represent sheaves localized on sub-manifolds of $X$.
The latter will be identified with the so-called fractional
brane states \rDM\rDFR. We will study in detail a
non-trivial example that demonstrates this appealing 
correspondence between the localization of two-dimensional fermions
and that of the associated fractional branes.

The multiplication rule of the chiral fields at the boundary leads
to a simple group-theoretical formula for the intersections of 
ground states in the LG phase, 
which coincides with the geometrical intersections
at large volume by a Hirzebruch-Riemann-Roch formula. This
correspondence between group-theoretical data of the 
orbifold singularity at small volume and the geometric intersections of 
the compact cohomology at large volume is a well-known problem in
mathematics, the McKay correspondence. We will identify the
bosonic/fermionic topological bases $\bR/\bS$ with the dual bases 
for the compact/non-compact K-theory group on $X$, introduced
by Ito and Nakajima in the case $\CC^3/\Gamma$. Thus the open string
point of view gives a completely natural explanation of the
McKay correspondence in terms of the interpolation between
the small and large volume phases of a two-dimensional 
quantum field theory, described by the group-theoretical 
structure of LG fields at small and exceptional
sheaves at large volume, respectively. 

The large volume representations
of the bases $\bR$ and $\bS$ carry an interesting mathematical
structure: they represent an exceptional collection of rigid
sheaves that give a foundation of a so-called helix of exceptional
sheaves \rRudi\rZas\rHIV. This structure comes with a distinguished operation
on the exceptional sheaves, the so-called mutation $\rmut$, and it
turns out that the topological bases $\bR$ and $\bS$ are 
related by a specific series of such mutations, 
namely $S^{a\, *}=\rmut^{N-a}\, R_a$. This relation is quite remarkable, 
as it provides an effective way to define the bases $\bS$ in terms 
of short exact sequences starting from the basis $\bR$.
One may then give yet another interesting representation 
of the topological bases  in terms of the local mirror description 
of exceptional sheaves derived by Hori, Iqbal and Vafa \rHIV. 
In the LG theory mirror, the two bases are identified as the unique 
complete, supersymmetric D-brane configurations and related by
a special monodromy.

Following the ideas of Diaconescu and Douglas \rDD, we 
may then use the D-branes on the K\"ahler manifold $X$ to study
D-branes on a generic, {\it compact} Calabi--Yau $Y$, by embedding
$Y$ in $X$ as a hypersurface. We will find a simple relation 
between the data on $X$ and $Y$ that allows us to define 
the topological bases $\bR$ and $\bS$ on the ambient space
or on the hypersurface interchangeably. This relation makes it possible
to work even on a {\it singular} ambient space if the hypersurface
avoids the singularities. This suggest a 
natural generalization of the McKay correspondence to 
singularities without a crepant resolution: as a relation between
the boundary ring of small volume LG fields and their large volume
sheaf counterparts on the smooth hypersurface of minimal 
codimension in the maximal crepant resolution of  $X$ provided
by the LSM.

The topological bases $\bR$ and $\bS$ provide also an extremely
effective description of the small/large volume 
D-branes on a generic, compact Calabi--Yau $Y$ in terms of the
simple, group theoretical data of the orbifold embedding space.
This improves substantially on the cumbersome closed string methods 
used so far, namely analytic continuation of periods \rDQ\ and 
the toric construction of refs.\rDGM\rDD. In particular we will show,
how a surprising amount of information on the
moduli space of $Y$, such as monodromy matrices and the analytic
continuation of periods, may be derived with ease 
from the topological bases of the open string sector. 

The organization of this paper is as follows: in sect. 2 
we summarize the structure of the zero mode sector of the
gauged linear sigma model with gauge group $H$ with an emphasis 
on the sector with boundaries coupled to gauge fields. 
In sect. 3 we argue that the boundary conditions select naturally
two bases of generators for the infinite dimensional zero mode
sector with boundaries. The first, $\bR$, is obtained from the trivial
ground-state by bosonic maps and is delocalized on the target
space. The second, $\bS$ arises from fermionic maps and localizes on
compact submanifolds. In sect. 4 we specify the two bases in
the small volume, orbifold phase and define an inner product
which is essentially the decomposition of tensor products
of $H$ representations in the multiplication of two-dimensional
chiral matter fields. The inner product is non-degenerate on the
two bases $\bR$ and $\bS$ and moreover the two are orthogonal
to each other.
In sect. 5, by a variation of the FI terms, we continue the
objects $R_a$ and $S^a$ through the moduli space and obtain 
their large volume definitions as 
geometric sheaves. Specifically, the elements of $\bR$ and 
$\bS$ are identified as exceptional sheaves that generate the infinite
space of sectors with boundaries in terms of complexes. We discuss also
some properties of a general pairing of the closed and open string 
sectors for target spaces with non-negative first Chern class.

In sect. 6 we connect the previous ideas to the McKay correspondence.
We identify the bosonic/fermionic bases with the generators of the
non-compact/compact K-theory groups on quotient singularities, that
have been introduced in the mathematical literature \rIN\rDD. 
In sect. 7 we argue that the bases $\bR$ and $\bS$ are foundations
of a helix structure and related by a series of mutations. 
This leads to a simple description of
the basis $\bS$ in terms of sequences involving the basis $\bR$.
Moreover, using the results of \rHIV, 
we identify the two bases as the unique complete, supersymmetric 
bases in the local mirror LG theory. 
In sect. 8 we turn to the case of Calabi--Yau hypersurfaces
embedded in a target space with $c_1>0$. We find a simple relation
between the K-theory data on the target space and the hypersurface
and use it to formulate a natural proposal for a McKay correspondence 
in singular resolutions.
Finally, in sect. 9 we apply the previous ideas to study D-branes 
on Calabi--Yau three-folds. We show that the intersection
matrix of the basis $\bS$ agrees with that derived from the Gepner model and
give an open string derivation of the topological data of the 
Calabi--Yau that enter the prepotential. On the base of an
explicit example we expose the correspondence between localization of
two-dimensional fermions and the fractional branes wrapped
on submanifolds in $Y$. Moreover we show how the D-brane spectrum, 
monodromies and analytic continuation matrices may be derived from the 
group theoretical data of the orbifold.

In the appendices we have collected a few simple examples to illustrate
various aspects of the discussion.

\newsec{Open strings and the gauged linear sigma model}
\subsec{The zero mode sector of closed and open strings}
We consider the two-dimensional $(2,2)$ supersymmetric gauge theory 
with gauge group $H$ and matter super-fields 
$X_i$ in representations of $H$ with an action of the form 
\eqn\eLSM{
S=S_{kin}+S_{gauge}+S_{FI}+S_W,
}
where the first two summands are the matter and gauge kinetic terms
and the remaining terms are the superpotentials for the twisted
and untwisted chiral fields, respectively. We refer to \rWlsm\rHV\rHIV\
for details and notation. In particular, \rHIV\ contains a profound
study of theories of this type in the presence of boundaries, with boundary
conditions on the fields that preserve 1/2 of the supersymmetry. For
related work, see \rOOY\rCYDbscft\rGJlsm.

The primary interest in these theories is
that the special class of conformal theories may describe the world-sheet 
theory of a type II string for an appropriate field content.
In this case, the boundaries correspond to BPS D-branes on which open strings
may end. However, as in \rHIV,  many of the following considerations 
make sense more generally, also for the non-conformal case and target
spaces of any dimension.
We will loosely refer to the sector with and without boundaries also
as the ``closed'' and ``open'' string sectors, respectively, though 
this nomenclature requires strictly speaking the above conditions to 
be satisfied.

An important example is the non-linear sigma model on an 
$n$-dimensional K\"ahler manifold $X$ with $W=0$ and no gauge fields. 
The lowest components  $x_i$ of the fields $X_i$ represent
coordinates on $X$ and the kinetic terms in \eLSM\ are determined
by the K\"ahler metric $g_{i\bb j}$ on $X$.
The left- and right-moving fermions $(\psi^i_+,\psi_+^{\bb i})$ 
and $(\psi^i_-,\psi^{\bb i}_-)$ in the super-fields $X_i$ are sections of 
$K^{1/2}\otimes (\Omega^{*\, (1,0)}\oplus \Omega^{*\, (0,1)})$ 
and $K^{1/2\, *}\otimes (\Omega^{*\, (1,0)}\oplus \Omega^{*\, (0,1)})$, respectively.
Here $K$ is the canonical bundle on the 2d world-volume $\Sigma$ and
$\Omega^{1,0}$ ($\Omega^{0,1}$) denotes the pull-back of the (anti-)holomorphic 
cotangent bundle of $X$ to $\Sigma$.

The zero mode sector of this sigma model without boundaries
has a topological nature. The ground states in the RR sector are
in one-to-one correspondence with the Dolbeaut cohomology on $X$ 
\rWindex\rRW.
The four supercharges act as the Dolbeaut operators $\p,\ \bb \p$ and 
their adjoints on the Hilbert space $\htopcl$ which can be
identified with the space of sections of 
\eqn\eclhs{\big(\wedge\,  \Omega^{(1,0)}\big)\, 
\wedge \,\big(\wedge\,  \Omega^{(0,1)}\big).}
By spectral flow the RR ground states are related to the so-called $(ac)$ 
ring of primary chiral fields in the NS$^2$ sector \rLVW. It is 
a deformation of the cohomology ring defined by wedging forms in $H^*(X)$. 
For this reason this ring is also called the quantum cohomology ring.
If $c_1(X)$ is zero, there is another correspondence
between the supersymmetric ground states and the elements of the
so-called $(cc)$-ring. We will mostly focus on the $(ac)$-ring, however.

The Hilbert space in the open string sector corresponds 
to the addition of boundaries and has again two sectors, 
denoted as the A-type and B-type boundary conditions, respectively
\rOOY\rHIV. They are naturally associated with the 
$(cc)$- and $(ac)$- ring of the
closed string sector. The 
boundary condition sets to zero the two linear combinations of the 
four fermionic zero modes that correspond to $dz^i$ and $i_{\p/\p z^i}$ \rHIV.
Moreover the fermions on the open string are coupled to 
the gauge field on the boundary D-brane. The topological 
sector of the open string Hilbert space
$\htopop$ is therefore identified with the space of sections of 
\eqn\eoshs{
\wedge\,  \Omega^{(0,1)} \otimes E^*_a \otimes E_b,}
where $E_a$ and $E_b$ are the gauge bundles that correspond to the
gauge fields on the two D-branes labeled by $a$ and $b$, on which 
the open string ends.
The two unbroken supercharges include the gauge fields
and act on the Hilbert space as 
the Dolbeaut operator $\bb\p_A=d\bb z^{\bb i}
(\p_{\bb i}+A_{\bb i}^{(b)}-A_{\bb i}^{(a)})$ and its adjoint,
respectively. 

Note that, contrary to \eclhs, the total space \eoshs, 
is infinite dimensional.

\subsec{Open strings in the gauged linear sigma model}
To discuss open strings we need to consider gauge fields.
We will argue that a minimal extension of the sigma model
by gauge fields, namely Witten's gauged linear sigma model construction 
of a K\"ahler manifold $X$ as a coset $G/H'$ \rWlsm, 
contains the necessary degrees of freedom to define a finite basis 
$B$ that generates $\htopop$, in a sense to be made precise.  

The gauge group $H$ of the gauged LSM is a subgroup of $H'$. The matter
fields $X_i$ carry representations of $H$ (and also of 
the global symmetry $G$) 
and their scalar components $x_i$ represent homogeneous coordinates on $X$.
Their fermionic super-partners $\psi^i$ take values in the 
tangent bundle of $X$, after taking into account the identifications 
made by the $H$ gauge invariance. 
Examples considered in \rWlsm\  include flag manifolds
with $H'=\prod_i U(n_i)$ and toric varieties with $H'= U(1)^r$. We will
discuss in detail mainly the case where the gauge group $H\subset H'$ 
is Abelian, though generalizations are possible and will be commented 
on along the way.

The ground states of the gauged LSM have a representation of the form 
\eqn\egsts{
f(x_i)\, \psi^{\al_1}\wedge\dots\wedge\psi^{\al_n},}
where $\psi^\al$ denotes any fermionic zero mode and $\al \in \{i,\bb i\}$. 
As the fields $X_i$ carry $H$ representations, the state corresponding to 
\egsts\ will be a section of some bundle $V$ which is determined
by the tensor product of the $H$ representations and the (anti-)holomorphic
tangent indices of the fermions. We identify $V$ as 
the ``difference bundle'' $E_a^*\times E_b$ of an open string sector
between boundaries carrying the gauge bundles $E_a$ and $E_b$, respectively.
In the closed string sector without 
boundaries, $V$ is trivial and the function $f(x_i)$ is determined, 
up to total derivatives, by the ordinary Dolbeaut operator. 
On the other hand, in the 
sector with boundaries, there will be an generically infinite number of 
allowed functions $f(x_i)$ corresponding to an infinite number of 
different $V$ valued Dolbeaut operators. 

Clearly we can generate only a subset $\htopopz\subset \htopop $ 
in this way, as the available bundles $V$ are constrained by the
representations of the fields $X_i$, and so are, loosely speaking,
a combination of $H$ bundles and the tangent bundle. More precisely,
the fermions contain the information not only about
bundles on $X$ but also sheaves supported on holomorphic submanifolds. 
The important point is that the bundles (or sheaves) that have a 
representation \egsts\ in the linear sigma model will turn out to
be sufficiently general to generate a finite basis $B$ for the 
infinite dimensional Hilbert space $\htopop$. In geometric
terms, the available sheaves will be generators for the derived 
category  $\derc(X)$, which means that any coherent sheaf 
on $X$ may be constructed in terms of (bounded) complexes of elements in $B$. 

\newsec{Two dual bases for $\htopop$ and localization properties}
As the space $\htopop$ \eoshs\ is infinite dimensional it 
is not obvious in the geometrical large volume phase what a 
finite basis for it might look like. The key point will be
the construction in the small volume  phase, which 
leads to a definite recipe for a construction of two dual, finite bases 
of generators with dimension $N=\dim H^{vert}(X)$, where $H^{vert}(X)=
\oplus_k H^{k,k}(X)$. 

Similar as we may build up the closed string sector $\htopcl$ from
the ground state $\bx 1$ by acting with the fermionic zero modes
on it, we start in the open string sector from the ground state
$\cx O$, corresponding to a section of the trivial bundle in 
the large volume limit. In fact the choice of the ``base point''
$\cx O$ is irrelevant and it may be replaced by any line bundle 
$\cx O(n_0)$, as the difference corresponds to a monodromy,
or, equivalently, to a change of the closed string background.

From \egsts, we expect that in some sense, we may obtain the sections
in $\htopop$ by multiplying sections of $\cx O$ with bosonic and
fermionic zero modes. The world-sheet point of view provides 
a more concrete approach. The B-type boundary conditions of the
$(2,2)$ supersymmetric sigma model (without $B$-field) are 
are \rOOY\rHIV\rNW\rGJS:
\eqn\ebcs{
\p_1\phi^\theta=\psi^\theta_--\psi^\theta_+=0,\qquad
\p_0\phi^n=\psi^n_-+\psi^n_+=0,}
where $\theta$ and $n$ are indices in the tangent and normal 
directions, respectively. The conserved supercharge is
\eqn\superc{
Q=\sqrt{2}\, \int 
g_{i\jb}\,\Big( 
(\psib ^{\jb} _+ + \psib ^{\jb} _-) \, \p_0\phi^i+
(\psib ^{\jb} _- - \psib ^{\jb} _+) \, \p_1\phi^i\Big).}
The representation theory of the supercharge $Q$
on the boundary has not been worked out but we will need only 
the following, simple observation. On the boundary, $Q$ splits into
the two parts
\eqn\supercii{\eqalign{
Q_{tangent}&=\sqrt{2}\, \int 
g_{\theta\bar{\theta}}\,
(\psib ^{\bar{\theta}} _+ + \psib ^{\bar{\theta}} _-) \, \p_0\phi^\theta,\cr
Q_{normal}&=\sqrt{2}\, \int 
g_{n\bar{n}}\,
(\psib ^{\bar{n}} _- - \psib ^{\bar{n}} _+) \, \p_1\phi^n.}}
It follows that the super-fields on the boundary have a 
structure similar to that of $(2,0)$ super-fields, with bosonic 
multiplets in the tangential and fermionic multiplets in the 
normal directions. Specifically, the components of 
the $(2,2)$ fields $\Phi^i=\phi^i+\sqrt{2}\theta^+\psi^i_++\sqrt{2}\theta^-
\psi^i_-+2\theta^+\theta^- F^i+\dots$ that survive the boundary conditions
may be assembled into super-fields of the form 
$\phi^\theta+\sqrt{2}\theta' \psi^\theta$
and $\psi^n+\sqrt{2} \theta' F^n$, respectively, 
with $\theta'$ the parameter for the surviving supersymmetry \superc.

We choose now the base point $\cx O$ in the 
infinite dimensional space \eoshs. As $\cx O$ is 
spread over all of $X$, the boundary
conditions are of Neumann type in all directions in this sector.
Multiplication of this state with the fields $\Phi^\theta$ yields 
another ground state with identical
spatial boundary conditions, but different gauge bundle $V$,
determined by the $H$ representation of $\Phi^\theta$.
On the other hand we may also consider a sector with a
new Dirichlet boundary condition on a submanifold $C$ in $X$.
In this sector, the lowest component of the LG field $\Phi^n$
is {\it fermionic} in the normal directions. Note that the fermionic
degrees of freedom are confined to the boundary in the
normal directions. Clearly this 
corresponds to a D-brane localized on $C$. 

We illustrate
this in Fig.1\fff for the type of geometry that we will focus
on in later sections, namely the resolution of a 
quotient singularity $\CC^n/\Gamma$. The ground state $\cx O$,
corresponding to the Neumann boundary condition in all directions,
projects onto bosonic super-fields. 
In the space-time sense, the D-brane that corresponds to
this sector has infinite mass due to the non-compactness
of the space. In the resolution with a compact exceptional 
divisor $C$, there are new boundary conditions that correspond to 
a finite mass D-brane on a compact cycle $C$. The projection of a $(2,2)$
multiplet at this boundary adds a fermionic super-field
that lives on $C$ and projects out the boson in the normal 
directions.

{\goodbreak\midinsert
\centerline{\epsfxsize 5.0truein \epsfbox{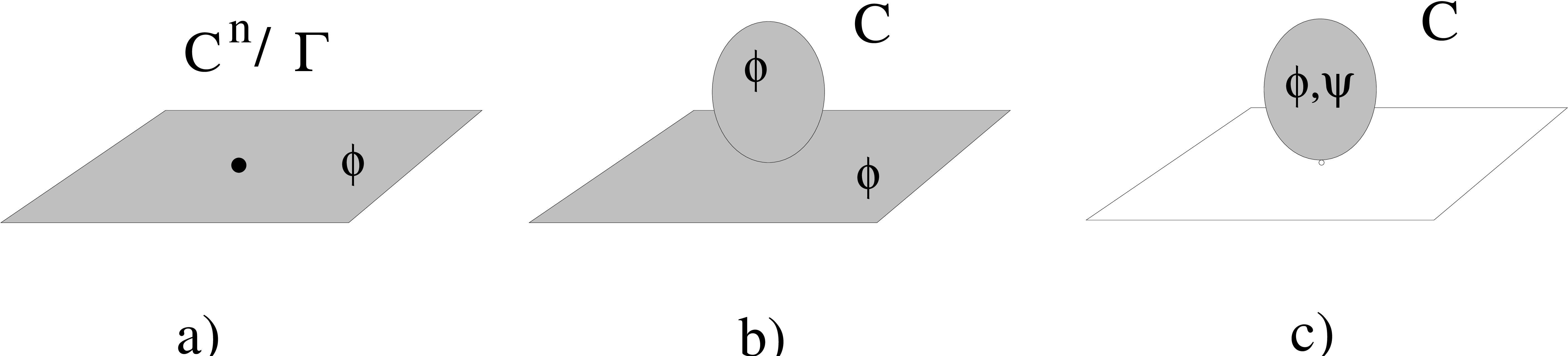}}
\leftskip 1pc\rightskip 1pc 
\noindent{\ninepoint  
{\bf Fig. 1\fff}: 
a) The ground state $\cx O$ for the trivial bundle on 
the non-compact space $\CC^n/\Gamma$. All directions are
tangential and the projection of a chiral multiplet at the boundary 
yields a bosonic super-field; $b)$ the resolution with 
compact exceptional divisor $C$ with the same boundary 
conditions; $c)$ the boundary condition that corresponds 
to a D-brane on $C$ projects onto fermionic super-fields
in the normal and bosonic super-fields in the tangential
directions.
}\endinsert}\vskip -0cm\ni
\ni
To construct bases of $\htopop$,
we may simply reverse the logic and note that multiplication of $\cx O$
by the lowest component of a $(2,2)$ super-field leads to a sector 
with a different bundle but the same spatial boundary conditions.
On the other hand multiplication by a fermionic zero mode $\psi^i$
corresponds to changing the boundary condition from Neumann to
Dirichlet in the directions normal to the hyperplane on which 
$\psi^i$ is localized. In the next section
we construct two bases $\bR$ and $\bS$ 
in the LG phase obtained by multiplication of $\cx O$ with only bosonic 
or only fermionic zero modes respectively. We will subsequently
study some remarkable properties of these bases and eventually 
show that they provide good finite bases of generators for $\htopop$. 
In agreement with the above localization arguments, 
the large volume version of the two bases $\bR$ and $\bS$ 
will correspond to bases for the general 
K-theory group $K(X)$, and the K-theory group $K_c(X)$ with 
compact support, respectively. This leads to a beautiful
identification of the bosonic/fermionic bases as dual bases of
a McKay correspondence, as introduced in the mathematical literature 
by Ito and Nakajima \rIN.

\newsec{Linear sigma model I: The group theoretical perspective}

Witten's gauged linear sigma model \rWlsm\rMPlsm\ description of
the K\"ahler manifold $X=G/H'$ is a 
(2,2) theory of the form \eLSM\ with canonical kinetic terms,
a gauge group $H\subset H'$ and matter fields $X_i$ in representations of $H$.
As will be reviewed in the next section, there are two distinguished
types of phases of this theories controlled by the FI terms,
corresponding to small and large volume, respectively. We will
first consider the small volume phase which
corresponds geometrically to some orbifold $\CC^{n+1}/\Gamma$, with
$\Gamma$ a discrete subgroup of $H$. It carries a natural 
group theoretical structure. In the presence of a superpotential
$W$, which we will add later, this phase describes a \LG theory.
We will first construct two bases $\bR$ and $\bS$ which are 
our candidates for a finite basis of generators for $\htopop(X)$ in this
phase. In the next section we carry the bases to large volume,
by varying the FI terms, and argue that the necessary conditions
for them to represent a basis of free generators are satisfied. 

\subsec{The case $H=U(1)$}
Let us start with the linear sigma model with gauge group $U(1)$ 
and $n+1$ matter fields $X_i$ of charges $w_i\in \ZZ$. As in \rWlsm\ 
it will be natural to extend the gauge group $U(1)$ to $\CC^*$. Its action 
on the scalar fields is given by $x_i\to \om^{w_i}\, x_i$, 
with $\om \in \CC^*$. If the weight vector $w=(w_1,\dots,w_{n+1})$ 
has only positive entries, $X$ is a compact weighted projective 
space $\WP^n_{w_i}$.

If the weights $w_i$ are all equal, this space is smooth and 
the above theory will also describe the geometric large volume
phase for appropriate values of the FI terms. Otherwise,
$X$ will have singularities at the fixed points of the $\CC^*$ action 
that have to be resolved to obtain a smooth space. The resolution
requires the addition of extra matter fields and $U(1)$ factors and
will be discussed in the next section.

\subsubsec{A basis $\bR$ from bosonic maps}
Let us first consider multiplication of the state $\cx O$ with 
the bosonic components of the LG fields, that is the homogeneous
coordinate ring.
As the fields $x_i$ carry only $U(1)$ charges, these states 
will flow to a basis of line bundles in the large volume phase. 
We denote a state with $U(1)$-charge $q$ obtained in this way $q$ 
by $\cx O(q)$:
\eqn\ecr{x_i:\cx O \to \cx O(w_i).}
As discussed already, we may shift the origin from $\cx O$ to $\cx O(-\infty)$.
In this way we obtain an infinite series of states with 
$U(1)$ charge $q\in \ZZ$. Let $\cx H_R=\{\cx O(q) \},\ q\in \ZZ$ 
denote this infinite set, 
ordered with increasing $U(1)$ charge.

\subsubsec{A dual basis $\bS$ from fermionic maps}
Instead of multiplying a ground state $\cx O(n_0)$ with $x_i$
we may consider, in view of \egsts, multiplication with 
the fermionic zero modes $\psi^i$:
\eqn\ecri{
\cx O(n_0)=S^1\matrix{\ss \psi[q]\cr\lra\cr\ }S^{q+1},}
where $\psi[q]$ is a product of fermions with $U(1)$ charge $q$.
We fix our conventions such that the
creation operator corresponds to a fermion that is a section of 
the tangent bundle; thus the objects $S^a$ live in the space dual to 
\eoshs. Different then before, the composition of fermionic maps 
is anti-symmetric and the construction terminates at charge $q=N$, which is 
the charge of the product of all fermions. In fact this 
combination of zero modes is equivalent to 0 by 
$U(1)$ gauge equivalence. Thus we get naturally a vector of $N$ 
elements $\cx E_S=\bS=\{S^1,\dots, S^{N}\}$. Note that there may 
be values $q=a'$, where no map \ecri\ exists and we may not construct the 
sector $S^{a'}$ with charge  $a'$ relative to $S^1$. These
ground states missing in the orbifold phase will be recovered
in the resolution of the orbifold, 
which introduces extra matter fields $Z_i$. 
In particular, in the resolution, the maps $S^1\to S^a$ exists for all 
values of $a$, with 
those missing in the orbifold phase provided by the fermions
in the new super-fields $Z_i$.

We can again define an infinite set $\cx H_S$ of bundles, 
consisting of an infinite number of copies of $\cx E_S$ 
with origin shifted by $N$ units of $U(1)$
charge. This is ordered set is dual to $\cx H_R$ w.r.t.
the bilinear product defined in the next section.
The fermionic zero modes carry also an index of the tangent bundle.
Therefore, the bundles in the large volume phase connected 
to the states in $\bS$ may have rank larger than one. 
The construction of these bundles in terms of sequences will
be discussed in sect. 5\fff, after we have described the smooth
resolution in the large volume phase.

\subsec{The Witten index}
The weighted number of closed string ground states \eclhs,
the Witten index $(-1)^F$ \rWindex, equals $\chi(X)$. The index in 
the open string sector $ab$ coincides, in virtue of \eoshs, with that of 
$\bb\p_{A}$, which, for a smooth space, 
is described by the  Hirzebruch-Riemann-Roch formula:
\eqn\HRR{
\rmx{ind}_{\bb\p_{A}}= \sum_k\, (-1)^k\, 
\rmx{dim}\, \Ext^k(E_a,E_b)\matrix{{\ss HRR}\cr=\cr{\phantom 1}}
\int_X \ch(E_a^*)\, \ch(E_b)\, \td(X).}
Motivated by the first expression, let us define an inner product 
$\la A,B \ra_H$ on elements in  $\htopopz$ as the
number of holomorphic maps $f$ from $A$ to $B$. 
For the ordered set $\cx H_R$, the maps $f_{a,b}$ are bosonic of 
degree $b-a$ and their number 
is equal to the number of independent monomials in the 
homogeneous coordinate ring with this degree.
For a reason that will become clear momentarily we
restrict to a basic set $\cx E_R$ of $N=\sum_i w_i$ consecutive elements 
in $\cx H_R$. Note that $N\cdot K$ is the first Chern class of $X$.
The index for the  elements $R_a\in \cx E_R$ is
\eqn\gisb{
\mchi^H_{ab} \equiv\la R_a,R_b \ra_H=
\Big(\prod_{i}\sum_{k=0}^{N-1} h^{k\, w_i}\Big)_{ab}=
\Big(\prod_{i}(1-h^{w_i})^{-1}\Big)_{ab}.}
Here $h$, is the $N\times N$ shift matrix with unit entries above 
the diagonal and zeros otherwise; it fulfills $h^N=0$. 
The formula \gisb\ contains only the group theoretical
information of $H$ and is, contrary to \HRR, well-defined even if
$X$ is singular. It will coincide with r.h.s. of  \HRR\ on a smooth
resolution $\tX \to X$ with the $R_a$ defined as the
appropriate pull backs\foot{This will be further discussed below.} to $\tX$.
We identify the degree $k$ with the fermion number of the map and thus
the only contribution to the index comes from $k=0$. 

Similar we may determine the inner product on the set $\cx E_S$,
where the maps carry non-trivial fermion numbers $k=0,\dots,N-1$.
The counting formula for these maps is the same as in the bosonic case,
up to an extra minus sign for each monomial $x_i$ and we obtain 
\eqn\gisf{
\chi^{H\, ab}=\la S^a,S^b \ra_H=\Big(\prod_i (1-h^{w_i})\Big)^{ab}.}
Note that the basis $\cx E_R=\bR$ from bosonic maps and the 
dual $\cx E^*_S=\{S^{a\, *}\}$ of the basis from fermionic maps are orthogonal 
with respect to the inner product $\la A,B \ra_H$, if the 
``base point'' matches. In fact it follows 
from \gisb,\gisf\ that the duals of the elements in the set
$\cx E_S=\{S^1,\dots,S^N=\cx O(-n_0)\}$ fulfill 
$\la S^{a\, *},R_b \ra _H = \delta ^a_b$. The $S^{a\, *}$ 
may be formally written as the linear combinations 
\eqn\sdef{S^{a\, *}=\chi^{H\, ab}R_b.}
In particular, eq.\sdef\ describes the relation between the Chern characters 
of the two  dual bases $\bR $ and $\bS$ on a smooth resolution $\tX$. 

\subsec{Generalizations to other gauge groups $H$}
The generalization of the above ideas for general $X=G/H'$ appears
to be relatively straightforward.
The matter fields $X_i$ come in representations $r_H$ of $H\subset H'$
and describe more general world volume gauge theories of D-branes wrapped 
on $X$. We may again define bosonic and fermionic maps, leading 
to representations generated by tensor products of $r_H$ and the conjugate
representations $\bb r_H$, respectively. 
The requirement that the elements in a basis $B$
represent free generators of $\derc(X)$ imposes a non-trivial 
selection rule on the allowed representations in $B$\foot{We will
formulate a conjecture for a group theoretical version of this
selection rule in sect.6\fff.}. The two dual bases $\bR$ 
and $\bS$ constructed in this way will again satisfy an orthogonality
relation $\la S^{a\, *},R_b \ra _H = \delta ^a_b$.

\newsec{Linear sigma model II: The geometric perspective}

\subsec{From small to large volume}
Consider the $H=U(1)$ theory 
with $n+1$ fields $X_i$ of positive charges $w_i$, now with 
one extra field $P$ of negative charge $-N$, where $N=\sum_i w_i$. 
As the sum of all charges is equal to zero, this a CFT. 
A supersymmetric vacuum of the theory must satisfy the D-term equations 
\eqn\dterms{
\sum_i\, w_i\, |x_i|^2-N\, |p|^2-r=0,}
where $r$ is the FI parameter of the $U(1)$. A variation
of the FI term interpolates between the geometric  and the LG Higgs phases
of the 2d QFT \rWlsm:

For positive $r$, at least one of the $x_i$ has to be non-zero.
The space parameterized by the $x_i$ divided by the $U(1)$ action 
(together with a careful treatment of the singular orbits \rWlsm\rMPlsm) gives 
the symplectic quotient construction of the weighted projective
space $X=\WP^n_{w_i}$. The scalar $p$ is a coordinate on the
bundle $\cx O_X(-c_1(X))$. The total space of
this bundle is an $n+1$ dimensional Calabi--Yau, non-compact in the
$p$-direction. This is the geometric Higgs phase where the 
$U(1)$ gauge symmetry is spontaneously broken by the vev's of the
$x_i$. The open string, or D-brane, states in this 
phase may be interpreted as elements of the K-theory on $X$ 
\ref\rRMa{Second ref. in \rDBcoups.}\rHaM\rWKth.

As $r$ is decreased to negative values, the size of $
X$ shrinks to zero (at least classically). From \dterms, we see
that the scalar field $p$ must be nonzero. As 
$p$ has charge $N$,
the $H$ gauge symmetry is broken by the vev of this field to 
the residual, discrete gauge symmetry $\Gamma=\ZZ_N$. There
are two different branches according to the values of the fields
$x_i$. At $x_i=0$, there is an unbroken gauge symmetry $\Gamma \in H$.
For $x_i\neq 0 $, also the subgroup $\Gamma$ is broken. As $p$
varies, the values of $x_i$ parametrize the geometric orbifold 
$\CC^{n+1}/\Gamma$.
This space is again a non-compact Calabi--Yau, with a quotient singularity 
at the origin. 

Similar as the variation of the FI parameter $r$ leads to a 
interpolation between LG and the geometric large volume phase
in the closed string sector, it connects the ground states of the
topological sector with boundaries in the two phases. Specifically,
the states $R_a$ and $S^a$ constructed previously are connected 
to sheaves in the large volume phase. In the following
we continue these states to  their large volume 
counterparts on the smooth resolution and argue that they 
generate $\htopop$. In particular in the large volume 
the multiplication with $x_i$ and $\psi^i$
is interpreted as a the multiplication ring of sections of \eoshs.

\subsec{Ring structures, a pairing and the finite basis of generators}
Before we proceed with an explicit construction of the finite
bases $\bR$ and $\bS$ of generators as bundles at large volume, 
let us discuss in which sense 
they will generate the infinite dimensional space $\htopop$.
Moreover we consider some interesting properties of 
a natural topological pairing with the closed string sector,
which determines, once again, the dimension of $\bR$ and $\bS$.

Let us assume for now that we work on a smooth space $ X$,
and the bases $\bR$ and $\bS$ are defined as sheaves on $ X$. 
The inner product $\gisb$ is non-degenerate, as is obvious from the
second expression. It follows that the basis $\bR$ generates the 
elements of the diagonal closed string Hilbert space $H^{vert}( X)$ 
by its Chern classes. The same is obviously true for the basis $\bS$.
This fixes in particular the dimension $N=\dim \, H^{vert}( X)$ of 
$\bR$ and $\bS$.

Similarly, the Chern classes of the 
bases $\bR$ or $\bS$ generate the Chern classes of 
all open string states by linear combinations.
A stronger requirement on a true basis $B$ of generators is that 
bounded complexes 
of elements of $B$ generate all sheaves on $ X$.
A necessary condition on $B$ is that there are
no higher Ext groups between the elements in $B$ and that
they provide free generators of the homotopy category 
$\derc( X)$ of finite complexes of sheaves on $  X$.
The first property is obvious for the basis $\bR$, as the only
contribution to the index \HRR, comes from $k=0$. It is in this sense that
the basis $\bR\subset\htopopz(   X)$ 
generates $\htopop( X)$.

The above is also true, though less obvious, for the basis $\bS$. The reason
is that the grading of the extension groups changes along the 
flow from small to large volume \rDFR\foot{%
It is also easy to see, that the flow does not change
the grading of the extension groups between the elements in $\bR$,
so that they remain a good basis also at large volume.}. This
will be further discussed on the basis of the explicit construction 
of the sheaves $S^a$.

Let us consider now an interesting topological pairing between the open and
closed string states provided by the Chern character.
For a  B-type boundary state $A\in H^p(X,V=E_a^*\otimes E_b)$ and 
a closed string state in the vertical cohomology $\eta\in H^{vert}(X)$,
we may consider the integral
\eqn\epairi{
( A,\eta ) = \int_{\tx\eta}\, \ch\, A,}
where $\tx \eta \in H_{k,k}(X)$ is the dual of $\eta$ and we use $A$ 
also to represent a section of the corresponding bundle. We could have
defined other parings that include non-trivial topological invariants of $X$
under the integral. Recall that the vertical cohomology 
$H^{vert}(X)$ comes with a distinguished, integral ring structure, 
namely the quantum cohomology ring \rLVW. Similarly, we expect a 
quantum ring structure to be defined on $\htopop$
along the lines of \rKon. We should therefore look for the 
distinguished pairing between the open and closed string Hilbert 
spaces that respects {\it integral} ring structures. 
Inspired by the integrality of the inner product \HRR, we consider 
its ``square root''
\eqn\eqdef{
Q^{(A)}=\ch(A)\sqrt{\td X}.}
This defines a charge $Q^{(A)}\in \cx H_{cl}$ which 
is the K-theoretic version \rDR\ of a formula obtained for the macroscopic 
RR-charge of a stringy D-brane by anomaly 
considerations \rDBcoups\rFW.
However \eqdef\ makes sense more generally
in the 
non-conformal (2,2) sigma model and with a target space of any dimension.
The charge $Q^{(A)}$ defines a specific pairing $(A, \eta )= 
\int_{\tx\eta} Q^{(A)}$. 
The index \HRR, rewritten in terms of $Q^{(A)}$ becomes
\eqn\eHRRii{
Q^{(A)}\cdot Q^{(B)}\equiv\la A,B \ra = 
\int_X \Big(\sum_k (-1)^k\, Q^{(A)}|_{k,k}\Big)\, Q^{(B)}\, e^{c_1(X)/2},}
where the subscript $|_{k,k}$ 
denotes taking the $(k,k)$-form part of an expression.
Note that all higher Chern classes of $X$ cancel out of this expression
so that it depends only on $c_1(X)$.

The bilinear form \eHRRii\ is defined on the 
infinite dimensional space $\htopop(X)$.
If $c_1(X)=0$, the expression \eHRRii\ is
symmetric (anti-symmetric) and displays
the orthogonal (symplectic) structure of the intersection form on 
the even (odd) dimensional Calabi--Yau $X$; it coincides
with it when restricted to a finite basis. In particular, for $n$ odd,
$\chi_{ab}$ becomes the Dirac-Zwanziger product in the 
space-time gauge theory obtained by a type IIA compactification on $X$.
For $c_1(X)\neq 0$, the form $Q^{(A)}\cdot Q^{(B)}$ still defines an
integral bilinear product on the Hilbert space $\htopop(X)$, but
has no symmetry properties\foot{In contrary,
the Dirac index 
$
\rmx{ind}_{\p \! \! \! \slash_A}=\int_X \ch(E_a^*)\, \ch(E_b)\, \hx A(X)
=\int_X \Big(\sum_k (-1)^k\, Q^{(a)}|_{k,k}\Big)\, Q^{(b)}
,$ with $\hx A(X) = e^{-c_1(X)/2}\, \td(X)$ the A-roof genus,
has good symmetry properties, but needs not to be integral in the 
given basis $B$. There is then an obstruction to define the
corresponding bundle on $X$.
The fact that the index $\HRR$ and the Dirac index differ is related
to the anomaly of the $(2,2)$ sigma model for $c_1(X)=0$, as
studied in detail in \rLW, see also \rFW\ for a related work.}.

We have not considered a relation between the $(cc)$ and A-type boundary 
states, which in the Calabi--Yau case provides a mirror description
of the above states on the mirror manifold.
A natural pairing between open and closed string
states has been studied in \rCVtat\rCVclass\rOOY\rHV\rHIV.
In a particular sector, it is related to the period integrals on $\tx X$.
It would be interesting to study the precise relation between this pairing 
and the one defined by \eqdef. We will comment on this connection
in more detail in sect. 7\fff.

\subsec{A basis of line bundles $\bR$ in the geometric phase}
If the weights $w_i$ are equal, $X$ is smooth and we can 
interpolate the states $\cx O(q)$ from small to large volume 
without further modifications. As the lowest bosonic components 
$x_i$ are sections of line bundles, the states $\cx O(q)$ will correspond to 
line bundles of Chern class $q\cdot K$ on $X$, with $K$ 
the hyperplane class of $X$. In particular eq.\ecr\ is naturally
interpreted as the multiplication of sections in $\htopop(X)$.
\foot{Recall the relation between line bundles 
and divisors. A line bundle $\cx L$ is characterized by
its first Chern class $[L]\in H^{1,1}(X,\ZZ)$, which is dual to 
an algebraic submanifold of codimension one, a divisor $L\in H_{n-1,n-1}(X)$. 
The divisor $L$ is locally defined as the zero of a meromorphic 
function $f$ in the homogeneous coordinates $x_i$. The first 
Chern class of the line bundle associated to $L$ is determined 
by the weights of $f$ w.r.t. to $r=h_{1,1}$ 
weight vectors $w^\al$ that represent the classes $H^{1,1}(X)$.
In fact $r$ is precisely the number of $U(1)$ factors in $H$ and 
the vectors $w^\al$ describe the charges of the matter fields $X_i$
under the gauge group $U(1)^r$. For the toric varieties we consider,
the line bundles are indeed in one-to-one correspondence with monomials in the
coordinates $x_i$,
as the toric divisors $D_i:x_i=0$ are known to span the Picard lattice
of $X$ \rBat.}

For general $w_i$, the $\CC^*$ action $x_i\to \om^{w_i}x_i$ has
fixed points that lead to singularities on $X$. To define 
the states $\cx O(q)$ properly as line bundles we need to 
specify their sections on a smooth resolution $\tx X$ of $X$. We proceed
to construct this bundles on a given (not necessarily unique)
resolution of $X$. Note that, independently of the chosen resolution,
the geometric inner product \HRR\ on the basis of bundles $\bR$ 
will coincide with the group theoretical formula \gisb. 

In the following we assume $n>1$ to avoid complications special to the 
low dimensional cases. We will also consider only a single resolution of $X$; 
additional resolutions may be treated step for step.
The linear sigma model in the new phase that corresponds 
to a partial resolution 
$\hx X$ has an $U(1)_{\hx X}^2$ gauge symmetry and one extra 
matter field $Z$. The size of the divisor introduced in 
this resolution is a new FI parameter $r'$.
The original $U(1)_X$ symmetry of the phase corresponding to $X$ is the linear 
combination of $U(1)^2_{\hx X}$ under which the field $Z$ is uncharged.

It is clear that 
any line bundle on $\hx X$  corresponds to a well-defined representation 
of $U(1)_X$; however the map from $q_{\hx X}$ to $q_{X}$ has the charge 
of the field $Z$ as its kernel.
To reconstruct the
basis $\bR$ on $\hx X$ from that on $X$ we require that {\it 
each map $R_a\to R_b$ generated by the field $x_i\in X_i$
in the phase $X$ pulls back to a map in the new phase $\hx X$}.
This determines uniquely the Chern class of the
bundle $\hx R_a$ on $\hx X$. 

E.g., if $R_N=\cx O$ and $x_0$ a charge one field in the LG
phase that provides a map $R_{N-1}\to R_N$, where $R_{N-1}=\cx O(-1)$
in the orbifold phase, then the Chern class of 
$R^{N-1}$ in the large volume phase is $c_1(R_{N-1})=-\sum^r_{i=1} 
q_0^\al\,  K_\al$. Here $q_0^\al$ are the charges of the field 
$X_0=(x_0,\dots)$ in this phase and the $K_\al$ the $(1,1)$ forms 
that correspond to the $U(1)^r$ symmetry of the LSM for the 
resolution $\tx X$. In other words, $\tx R_{N-1}=\cx O(-q_0^1,\dots,-q_0^r)$,
and similarly for the other bundles $\tx R_a$.
For an explicit example, we refer to sect. 9.3\fff.

\subsec{The dual basis $\bS$}
From the previous considerations it is rather evident what kind of 
objects the $S^a$ are in the geometric phase: the sheaves
of sections generated by multiplication of a section of the
line bundle $\cx O(n_0)$ with fermionic zero modes. The massless
fermions of the LSM are described by the exact sequence:
\eqn\fermo{
0\lra \cx O^r \lra \oplus_{i=1}^{n+r} 
\cx O(q^1_i,\dots,q_i^r) \lra \Omega^* \lra 0,}
where $H=U(1)^r$ is the gauge group of the LSM on the
resolution $\tx X$ and $q_i^\al$ the charges of the 
$n+r$ matter fields $X_i$ and $Z_i$. 
The above is just the 
statement that their fermionic components are sections of the 
tangent bundle of $\tx X$ and carry 
$U(1)^r$ charge $q_i^\al$.

In particular, if the weights are equal, $w_i=1$, 
then the sheaves $S^a$ are, by construction, simply the $a$-th exterior power 
of the twisted tangent bundle
\eqn\espow{
S^{a}=(-)^{N-a}\, \Lambda^{a-1} \Omega^*\otimes \cx O(-n_0-a).}
Here we have included a minus sign from the fermion number of 
the map in the definition. Note that $S^{N}=\Lambda^{N-1}\Omega^*(-n_0-N)$ 
is the line bundle $\cx O(-n_0)$; the dual basis $\{S^{a\, *}\}$
is thus orthogonal to the basis of line bundles
with $R_N=\cx O(n_0)$ in the orbifold phase.
By construction, the exact sequence associated to the 
bundles $S^a$ is simply the appropriate exterior power of \fermo !

In general, if the $w_i$ are not equal, the structure is similar
and may be inferred from the two-dimensional point of view
described in sect. 3\fff. Roughly speaking, {\it the objects 
$S^a$ represent exterior powers of the tangent bundles on
compact, holomorphic submanifolds in $\tx X$}, including $\tx X$ itself.

As discussed above, the resolution $\tx X$ of the singularities of $X$
introduces new matter super-fields $Z_i,\ i=1,\dots,r-1$ together
with $r-1$ new $U(1)$ gauge multiplets. The former are associated with
the exceptional divisors of the resolution defined by $D_i:\, z_i=0$.
Moreover, their fermionic super-partners $\zeta^i$ generate 
new Dirichlet boundary conditions along $D_i$ and 
intersections thereof. 

It is relatively straightforward to proceed from this general arguments 
to a more detailed description of the sheaves $S^a$ in a concrete example.
In sect. 9.3\fff we will study in detail a representative case 
that demonstrates the nice correspondence between the localization 
of the two-dimensional fermions
and that of the fractional branes  $S^a$, 
in complete agreement  with the above picture\foot{Another 
instructive example is given in appendix B.}. Moreover, 
we will find a convenient closed form for the sequences 
that describe the $S^a$ in sect. 7\fff in terms of mutations 
of exceptional collections. Here we restrict to outline 
the general structure from the two-dimensional world-sheet point
of view. 

We choose $S^1$ to be the trivial bundle $\cx O$ and assume that there
are $k+1$ fermions $\psi^i$ of charge 1. Then the  sheaf $S^{2}$ 
is defined by the sequence
\eqn\exsi{
0 \lra \cx O \lra \CC^{k+1}\otimes \cx O(1) \lra S^2 \lra 0,}
and it has rank $k$. Here $\cx O(a)$ denotes the $a-$th power of
the hyperplane bundle and we have assumed that we may eliminate the
extra matter fields $Z_i$ introduced in the resolution in this step.
The bundle $S^1$ describes the pull back 
of $\Omega^*_{E}$ to $X$, where $E$ is a smooth, holomorphic submanifold 
of dimension $k$ in $X$, parametrized by the
bosonic super-partners of the $\psi^i$.
If $k>2$, then $S^3$ is given by the sequence
\eqn\exsii{
0 \lra S^2 \lra \CC^{(k+1)k/2}\otimes \cx O(1) \lra S^3 \lra 0.}
Thus $S^3$ is the pull back of the second exterior power $\wedge^2
\Omega^*_{E}$, and so on. 
At the $k+1$-th step this procedure terminates as 
$\wedge^{k+1}\Omega^*_{E}$ does
not exist. At this point there will be a new set of 
fermions of charge $k+1$. 
If this set contains again some of the fermions $\psi^i$ that are already
present in the orbifold phase, the series of bundles $S^a$ continues with 
another set of pull backs of (possibly twisted) tangent bundles on 
a submanifold in $X$. 

A new situation arises if at some point 
there is no such fermion $\psi^i$, or it is equivalent to zero by the
gauge invariance. In particular this happens if
there is no map $\psi[a]:S^1\to S^{a}$ 
in the orbifold phase. In the resolution there will then be 
an additional fermion $\zeta^i$ of the appropriate charge.
It is the fermionic component of one of the extra fields $Z_i$ and the Dirichlet
boundary condition imposed by it sets the bosonic component $z_i$
to zero. The submanifold in $X$ defined by this zero is an
exceptional divisor $D_i:\, z_i=0$. As the 
sheaf $S^a$ does not involve other fermions that live on $\tx X$, 
there are no non-zero sections away from the divisor $z_i=0$.
In other words,  $S^a$ is a sheaf on $D_i$, extended by zero on $X$. 
As the codimension of $D_i$ is one, there is precisely one such 
fermion and the sheaf $S^a$ is in fact the extension by zero 
of the line bundle  $\cx O(q^\al_{Z_i})_{D_i}$.

Note that this is the same argument that provides also the reason 
for why the basis $\bS$ is localized on the compact exceptional divisor $X$ 
of the partial resolution $\cx O(-c_1(X))_{X}$ of $\CC^{n+1}/\Gamma$. 
In this case, 
the relevant fermion that imposes the Dirichlet boundary condition 
for the resolution is the super-partner of 
the coordinate $p$ on the fiber.

\newsec{The McKay correspondence}
In the previous section we constructed two finite bases of generators for 
$\htopop$,
one from maps build from bosonic fields and the other 
from fermionic fields, and related them to sheaves in 
the geometric large volume phase. The intersection form
of the two bases is determined by the group theoretical
formulae \gisb,\gisf\ in the orbifold phase and coincides
with the geometric formula \HRR\ in the resolved phase. 
Moreover, by the non-degeneracy of the inner product $\chi_{ab}$,
they generate the closed string Hilbert space 
by their Chern classes if $\dim H^{vert}(\tx X)=N$.
More precisely, the bases $\bR$ and $\bS$ are two bases of 
generators for the topological K-Theory group $K(\tx X)$,
if $H^*(\tx X)=H^{vert}(\tx X)$. This is true for the case of toric 
manifolds that we consider \rBD.

A relation between the group theoretical data of the orbifold 
$\CC^{n+1}/\Gamma$, and the homology of its resolution is a well-known
subject in mathematics, the McKay correspondence. In particular 
Ito and Nakajima have introduced two bases for K-theory groups
on $\CC^3/\Gamma$ to formulate and prove a McKay correspondence
in this case \rIN\foot{The importance of these bases for D-branes
on the orbifold has been emphasized in \rDM\rDD.}. 

From the previous considerations we see that {\it 
the McKay correspondence
has a completely natural explanation in terms of the 
$(2,2)$ supersymmetric sigma model with boundaries}.
As the connection with our study of the open string Hilbert space 
in the previous sections is rather clear, we will be brief 
in the following. 

\ni
{\bf Claim 1} {\it The continuation of the 
bosonic and fermionic bases $\bR$ and $\bSd$ of $\htopop$ 
to large volume provides two orthogonal bases for the compact
K-theory group $K_c(\tx X)$ on the linear sigma model resolution 
of $\CC^n/\Gamma$. Moreover 
the basis $\bR$ extends to a basis of the K-theory
on the non-compact space $\CC^{n+1}/\Gamma$. } 

\ni
In particular, for $\CC^3/\ZZ_N$, 
we identify the bases $\bR$ and $\bSd$ with 
the restriction of the constructions of Ito and Nakajima to 
the compact part $\tx X$ of the resolution. This is clear 
from the fact that $i)$ $\bR$ is a set of line bundles that 
generates $K(\tx X)$ $ii)$ it provides a complete set of irreps of $\Gamma$
(see below) $iii)$ the basis $\bSd$ is  orthogonal to $\bR$.
Note that the open string point of view gives
a literal identification between tensor products of $\Gamma$
irreps and intersections on the resolution, in terms of the
HRR identity \HRR\ and its group theoretical form \gisb,\gisf.
More specifically the interpolation via variation of FI terms
gives also a continuation between the objects in the two
different phases, LG fields in $H$ representations in the 
orbifold phase and sheaves in the large volume phase, respectively.

A point that needs a little explanation is that the
formula \gisf\ involves representations of 
$H$ and not of the discrete group $\Gamma$. However this issue is
clear from the discussion in sect. 5.1: in fact $\Gamma$ is 
the unbroken part of the continuous gauge group $H$ in the
small volume phase, and the tensor products of the $H$ representations
descend to those of $\Gamma$.

For this to be true it is of course necessary that the 
map from $H$ to $\Gamma$ reps is injective and onto irreps.
In the present case of $H=U(1)^r$ this is obvious, by construction.
In fact we have reconstructed the bases on the resolution $\tX $
from the small volume basis $\bR=\{\cx O(n_0-N+1) ,\dots,\cx O(n_0)\}$.
The elements of $\bR$ define a complete set of
irreducible representations $\rho_a$ of $\Gamma=\ZZ_N$ by the
identification
\eqn\erid{
R_a \lra \rho_\al,\qquad \al = a\, \rmx{mod}\, N,}
where $\rho_\al$ transforms under the $\ZZ_N$ with eigenvalue
$\om^\al$. Here $\om$ is an $N$-th root of 1, $\om^N=1$.

For general gauge group $H$, the requirement that the map from $H$
to $\Gamma$ reps is injective and onto irreps may imply a
non-trivial selection rule on $H$ representations. This 
motivates the following conjecture:

\ni
{\it Conjecture}: { Consider a the set of generators 
$\{\tx S^{a}\} \subset \htopop(\tx X)$
constructed as in the previous sections.
Then there exists a subset $\bS$ of $N$ elements 
in $\{\tx S^{a}\}$ such that the two following conditions are
equivalent:
\item{$i)$} upon restriction from $H$ to $\Gamma$, $\bS$ provides
a complete set of irreps of $\Gamma$. \item{$ii)$} the subset $\bS$ 
provides free generators of the derived category $\derc(\tx X)$.}

\newsec{Helices, mutations and the local mirror description}

By construction, the two bases of generators $\bR$ and $\bS$
are defined in the orbifold limit as a set of $N$ consequent elements 
of the infinite sets $\cx H_R$ and $\cx H_S$, respectively, as 
defined in sect. 4. We will now study this structure in more detail
and identify the basis $\bR$ as the foundation of a helix $\cx H_R$
of exceptional sheaves, and similarly for $\bS$ and $\cx H_S$. In
particular this will lead to a very effective definition of the
basis $\bS$, in terms of short exact sequences involving the 
elements of the dual basis $\bR$.

We will also consider a different relation between
sheaves with a large volume interpretation
and states of a $(2,2)$ supersymmetric orbifold, namely 
local mirror symmetry. In a profound study of boundary states 
in $(2,2)$ supersymmetric two-dimensional QFT's\rHIV, 
Hori, Iqbal and Vafa showed that helices of 
exceptional sheaves in a linear sigma model on a Fano variety $X$
are related by local mirror symmetry to A-type boundary states
of a \LG theory\foot{Although the discussion
in \rHIV\ involves a certain class of examples, the
arguments apply more generally and have a straightforward generalization
to many other spaces.}. The latter correspond 
to D-branes wrapped on Lagrangian cycles in the mirror manifold.

In particular, the local mirror analysis of \rHIV\ was phrased 
in the mathematical framework of helices of exceptional sheaves.
The link between
helices and LG theory had been observed some time 
ago \ref\rKonii{M. Kontsevich, as quoted in \rHIV.}\rZas\
and its explanation has been one of the subjects of ref.\rHIV.
Another relevant relation is the one between 
helices and quivers as an equivalence between derived 
categories \rBon.

\subsec{A helix point of view}
Let us sketch the definition of a helix of coherent 
sheaves; for more details we refer to \rRudi\rZas.
An {\it exceptional sheaf} $E$ has $\Ext^0(E,E)=\CC$ and $\Ext^k(E,E)=0$
for $k>0$.
An {\it exceptional collection} of sheaves is an ordered collection 
of exceptional sheaves $\cx E=\{E_1,\dots,E_N\}$ such that 
$\Ext^k(E_a,E_b)=0$ for $a>b$ and for $a<b$ except at most for a single 
degree $k=k_0$. In particular,
the index \HRR\ has at most one non-trivial term equal to
$(-1)^{k_0}\Ext^{k_0}(E_a,E_b)$. 
A {\it strong exceptional collection} has $k_0=0$ and may be
used as a starting point to construct a 
map from the derived category of coherent sheaves 
$E_n$ to the derived category of quiver algebras with relations
\rBon.

The exceptional collection $\cx E$ has the important property one may
possibly define an 
operation, called {\it mutation} that acts on a pair in $\cx E$
and yields another exceptional set. One distinguishes left and right
mutations that act as $\lmut:(E_{a-1},E_{a})\to (\lmut_{a-1}E_a,E_{a-1})$
and $\rmut:(E_{a},E_{a+1})\to (E_{a+1},\rmut_{a+1}E_a)$, respectively.
The Chern class of the mutated sheaves have a simple form,
up to sign:
\eqn\mutchc{\eqalign{
|\  \ch{\lmut_{a-1}E_a} \ |=\ch{E_a}-\chi(E_{a-1},E_a)\ch{E_{a-1}},\cr
| \ \ch{\rmut_{a+1}E_a}\ |=\ch{E_a}-\chi(E_{a},E_{a+1})\ch{E_{a+1}},}}
where the sign depends on the details (and represents
the orientation in the context of D-branes \rHIV). 
Moreover the operations $\lmut$ and $\rmut$ satisfy 
the braid group relations. The definition of the operations 
$\lmut$ and $\rmut$ depends on details of the elements in 
$\cx E$. We refer to \rRudi\rHIV\ for a list of the definitions in the
various cases.

Finally, the {\it helix} is defined as an infinite collection $\{E_a\}$
such that the set $\{E_{n_0+1},$$\dots,E_{n_0+N}\}$ is an exceptional
collection for any $n_0$. Moreover $\lmut^{N-1}E_{n_0+N}=E_{n_0}$ and
$\rmut^{N-1}E_{n_0+1}=E_{n_0+N+1}$, that is the $N-1$-th 
powers of $\lmut$ and $\rmut$ generate shifts of the origin by
one to the left and to the right, respectively. In this way one
may obtain the infinitely extended helix starting from an exceptional
collection, which is then referred to as the foundation of the helix.
The integer $N$ is called the period of the helix.

We observe that the set $\cx H_R$ defined in sect. 4\fff obeys the
criteria of the definition of a helix.
Its period is defined by the first Chern class $c_1(\tx X)$
of the exceptional divisor. The basis $\cx E_R=\bR\subset \cx H_R$
of $K(\tx X)$ provides a foundation of the helix $\cx H_R$.

An interesting question is what the meaning of the
mutations is in terms of the McKay bases $\{R^a\}$ and $\{S^a\}$.
An evident operation is $\lmut^{N-1}$ ($\rmut^{N-1}$) 
which represents shift  of the origin by one to the left (right)
and leads to a foundation that generates the same helix.
An arbitrary mutation leads in general  to a foundation 
which generates a different helix. By \mutchc\ the classes of the new basis
$\tx {\cx E}$ present a linearly independent set of combinations of the 
classes in $\cx E$.

\subsec{Mutations in the LSM and a generator for the helix $\cx H_R$ }
In \rHIV, mutations have been identified as the monodromies in the
moduli space of the local mirror LG model. However it is important
to note that the moduli space of the LSM provides only an extremely
restricted set of perturbations, and only a very small subset of
mutations will be realized as monodromies. A canonical 
set of monodromies associated with the large volume limit of the LSM
are the shifts $t_\al \to t_\al +1$. They correspond to a shift of the 
$B$-field by $K_\al$. 

In sect. 9.3\fff we argue that there are always two distinguished 
monodromies, one, denoted by $A$,  around the orbifold point and another,
denoted by $T$, around a
divisor in moduli space where a D6 brane shrinks to zero size.
Moreover the inverse $T_\infty^{-1}$ of the combined monodromy $T_\infty=AT$ 
corresponds to the series of mutations
\eqn\shiftmut{
\{R_1,\dots,R_N\}\lra \{R_2,\dots,R_{N+1}\},}
and generates the helix $\cx H_R$ from the foundation $\bR$. 
Note that we may use the monodromy $T_\infty$ to generate also the
foundation $\bR$ from a single element, say $R_1$. 
In the orbifold phase, the analogue operation that generates the 
infinite set of ground states $\cx H_R$ is the multiplication
\ecr\ by a LG field of charge 1. Thus $T_\infty$ is the
generalization of this generating element to the resolved phase
of the LSM. It is interesting to observe, that 
this monodromy is not an element of the large volume monodromy.
It also does not preserve the distinguished large volume limit in general,
see sect. 9.3\fff for an example. 
This is another reflection of the fact that
the objects $\bR$ are well-defined throughout the moduli space,
which makes the notion of a mutation meaningful even through the 
non-geometric regions in the moduli.

\subsec{A reflection mutation $R_a\leftrightarrow S^{a\, *}$}
We will now consider an important series of mutations that does not 
correspond to a monodromy of the linear sigma model. Its special 
feature is that it maps the dual bases $\bR$ and $\bSd$ onto
each other. Specifically, we consider the 
operation $\pmut$ that acts as a reflection 
on a foundation $\cx E$:
\eqn\defp{
\pmut:\ \cx E=\{E_1,E_2,\dots,E_N\} \longrightarrow 
\tx{\cx E}=\{E_N,\rmut E_{N-1},\dots,\rmut^{N-1}E_1\}.}
From \mutchc\ we find that the Chern class of $\tx E_a$ is
\eqn\chcl{
\tx{E_a}= E_a+\sum_{b=a+1}^N c_{a,b} E_b,}
where we write simply $E_a$ for $\rmx{ch}(E_a)$ and 
\eqn\defc{
c_{a,b}=-\chi_{ab}+\sum_{l=2}^{b-a}\  (-)^l \hskip -1cm
\sum _{_{\qquad \ a<n_1<\dots<n_{l-1}<b}}
\hskip -1cm \chi_{a,n_1}\, \chi_{n_1,n_2}\, \dots \chi_{n_{l-1},b}.}
These equations can be rewritten in compact form as 
\eqn\sdefm{
\tx E_{N-a} = \Big( \sum_{l=0}^N (-)^l \tx \chi^l \Big)^{ab}\, E_b
= \Big( \fc{1}{1+\tx \chi} \Big)^{ab}\,  E_b = \Big(\chi^{-1}\Big)^{ab}\, E_b,}
where $\tx \chi = \chi -1$ satisfies $\tx \chi^N=0$. 
In particular, if we set $E_a=R_a$, we may identify 
\eqn\sform{
S^{a\, *}=\rmut^{N-a}\, R_a.}
We see that the basis $\bSd$ of $K_c(X)$  represents
the (dual of the) mutation $\pmut$ of $\{R^a\}$ of $K(X)$ and moreover 
$\tx{\cx E_S}=\{S^1,\dots,S^N\}=\{S^{N\, *},\dots,S^{1\, *}\}^*$ 
is the foundation of a
helix.

\subsec{A monodromy interpretation}
In ref.\rHIV\ it was shown, that the exceptional bundles
$R_a$ are mapped to A-type boundary states of a LG model by
local mirror symmetry. Moreover mutations of the exceptional collection
have been identified with
the monodromies of the LG model under variation of its deformation
parameters. It is instructive to consider the monodromy corresponding
to $\pmut$. This is best illustrated in the $W$-plane, the complex plane 
corresponding to the values of the superpotential in the mirror LG
model:
{\goodbreak\midinsert
\centerline{\epsfxsize 4.0truein \epsfbox{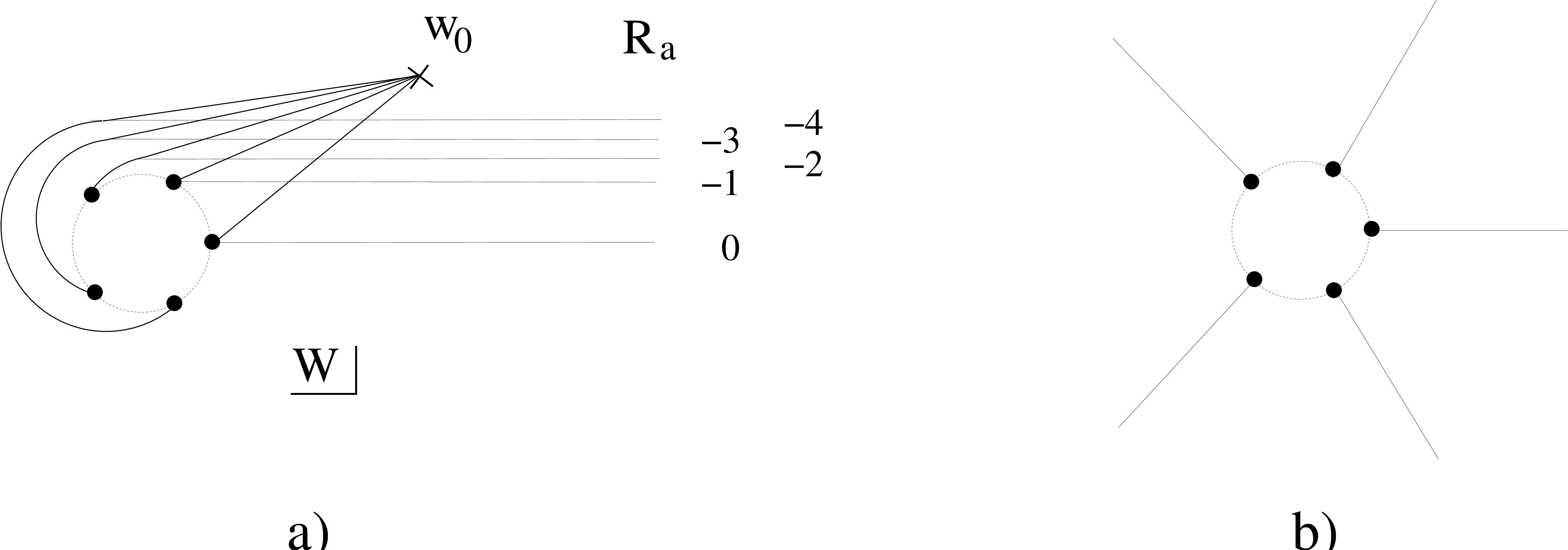}}
\leftskip 1pc\rightskip 1pc 
\noindent{\ninepoint  
\centerline{{\bf Fig. 2\fff}}
}\endinsert}\vskip -0cm\ni
The five circles in Fig.2a)\fff correspond to the 
$\ZZ_5$ symmetric vacua $w_a$ of 
the mirror LG model of $X=\IP^4$. By a standard
construction in singularity theory, one may define a complete
basis of Lagrangian 4-cycles\foot{See sect. 2 of \rHIV\ for a review
of this construction.} in the space of LG fields by choosing
a point $p$ and considering its preimage $W^{-1}(p)$.
This definition involves a choice of paths $\gamma_a$ from $p$ to the 
critical points $w_a$ that defines a choice of basis;
a change of the homotopy class of
$\gamma_a$ by moving it through a critical point $w_b$ corresponds to
a monodromy on the 4-cycles. By aligning the 4-cycles along the path 
$\gamma_a$ and moving $p$ to infinity, one may similarly define a 
basis of non-compact 5-cycles $C_a$ spanning $H_5(\CC^5,B)$, the homology 
of cycles with boundaries on the boundary $B:\ |W|=\rmx{const.}$
Moreover the paths $\gamma_a$ are identified in \rHIV\ as the
image in the $W$-plane of the D-branes
wrapped on the cycles $C_a$. The supersymmetric D-brane wrappings
correspond to straight lines in the $W$ planes with the 
slope determined by the phase in the definition of the preserved supercharge.
This is shown in Fig 2b)\fff.

The choice of path in Fig.2 corresponds to the mirror image of the 
exceptional collection $\cx E=\{R_a\}=\{\cx O(-4),\dots, \cx O\}$ \rHIV.
As indicated, an ordering is defined by the vertical coordinate of a line
to infinity, with an obvious modification if $p$ is at a finite value.

A right (left) mutation of $E_a$ corresponds to a monodromy in the 
$W$-plane, where the $a$-th line moves through the critical value
next to it in (counter-)clockwise direction. The choice of path
generated by the action of the reflection $\pmut$ is shown
in Fig. 3a\fff.
{ \sl
\goodbreak\midinsert
\centerline{\epsfxsize 4.8truein \epsfbox{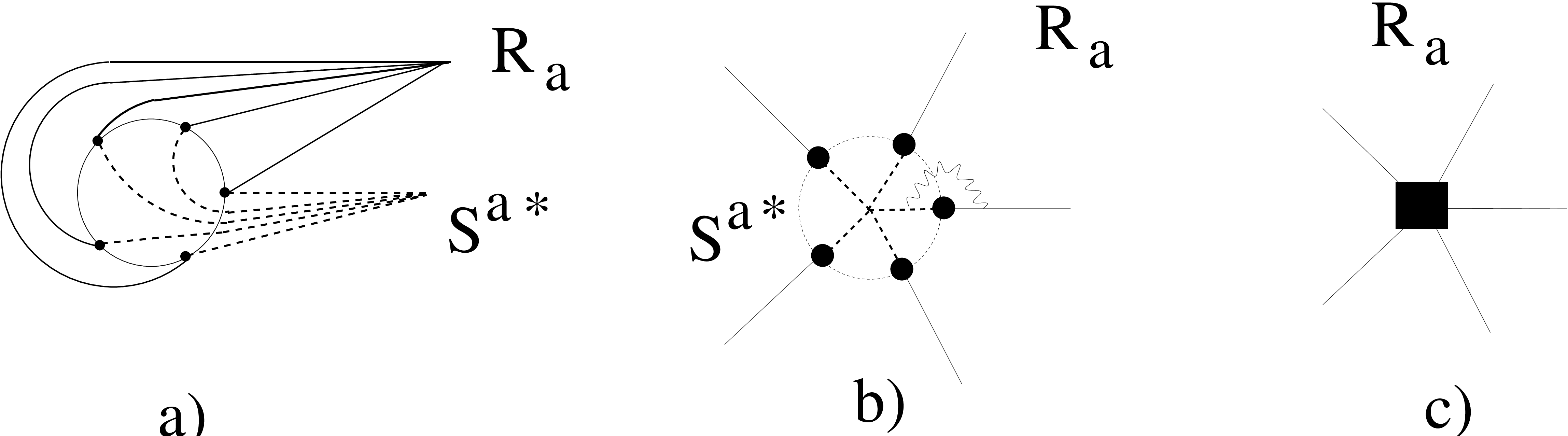}}
\leftskip 1pc\rightskip 1pc 
\noindent{\ninepoint  \baselineskip=8pt 
{{\bf Fig. 3\fff: the mirror of the two McKay bases :} 
$a)$ the two McKay bases $\bR$ and $\bSd$ 
are related by the monodromy $\pmut$.\
$b)$ the supersymmetric brane
configurations converge to the points $\bx \infty$ and $\bx 0$, respectively.\
$c)$ in the orbifold limit, the supersymmetric cycles in $\bSd$ collapse,
while the non-compact basis $\bR$ survives.}
}\endinsert}\vskip-0cm
The bases of {\it supersymmetric} cycles homotopic to those in Fig. 3a\fff
is shown in Fig. 3b)\fff. We see that the bases $\bR$ and $\bSd$ correspond
to the two unique complete bases of supersymmetric D-branes
in the \LG model. The orthogonality property of the two bases is evident,
as the single possible massless ground state between 
the brane $R_a$ and the brane $S^{b\, *}$ is an open string 
sitting at the critical point $w_a$. 

In the orbifold limit, the critical points $w_a$ move to the 
origin as shown in Fig 3c)\fff. The basis $R_a$ of non-compact 
cycles survives, while all the supersymmetric cycles in $\bSd$ 
collapse to a point. 

\subsec{A simple series of sequences for the basis $\bS$}
The monodromy relation \sform\ between the bases $\bR$ and $\bSd$ 
leads to a simple construction of the sheaves
$S^{a}$ in terms of mutations. The definition of the mutation
depends on the properties of the foundation $\cx E$; we 
refer to \rBon\rHIV\rZas\ for the definition in the various cases.
We will write here the sequences for the dual basis $\bSd$
which is orthogonal to $\bR$. The sequences for $S^a$ are simply the
transpositions of those for $S^{a\, *}$.
For the set $\bR$ a right mutation may be defined by the exact sequence\rRudi:
\eqn\rmutdef{
0 \lra E_a \lrai{\al}
\Ext^0(E_a,E_{a+1})\otimes E_{a+1}\lrai{\be} \rmut E_{a}\lra 0.}
E.g. if the $w_i$ are equal and thus $X=\IP^n$, this is a  
twisted form of the well-known Euler sequence
\eqn\eulers{
0\lra \cx O(k_a) \lra \CC^{n+1}\otimes \cx O(k_a+1)\lra \Omega^*(k_a) \lra 0,}
where $k_a=n_0-N+a$ for the elements of the 
foundation $\bR$ with $R_N=\cx O(n_0)$. 
Similarly, from the relation $S^{a\, *}=\rmut^{N-a}\, R_a$, 
we may obtain the sequence for the sheaves $S^{a\, *}$ by 
repeated application of the mutation \rmutdef. It is simply the
appropriate exterior product of the sequence \eulers, as can be
easily seen by explicit construction of the maps. In this way
we recover the result
\eqn\sfqeom{
 S^{a\, *}=(-1)^{N-a}\, \Lambda^{a-1} \Omega(n_0+a),\ a=1,\dots,N.}
Specifically, eq.\sfqeom\ is the dual of \espow.
Although the argument of multiplication by fermionic zero modes 
leads much more directly to this identification, 
the language of mutations provides a convenient closed
form of the involved sequences for arbitrary weights $w_i$. 
On the other hand it is not guaranteed that it is always possible to define 
the required series of mutations, as the elements of the mutated 
foundation may have Ext groups different from that in the original one. 
In this case, the two-dimensional point of view in
sect. 5.4\fff provides a more 
general framework for the definition of the appropriate sequences.

\newsec{D-branes on compact Calabi--Yau's and a 
McKay correspondence for singular resolutions}

So far we have defined the sheaves $R_a$ and $S^a$ on the exceptional
divisor $X$ of a partial resolution of the singularity $\CC^{n+1}/\Gamma$.
We are now, as in \rDD,  interested to study D-branes on a generic,
compact Calabi--Yau $Y$, embedded as a hypersurface in 
$X$. This corresponds to the addition of a homogeneous
superpotential $W\neq 0$ to the $(2,2)$ supersymmetric 
field theory \eLSM.

The main result of this section 
is a simple relation between the data on $X$ and $Y$
that allows to go forth and back between the definition of
the bases $\bR$ and $\bS$ on $X$ and their restrictions to
$Y$. This will allow us to define $\bS$ and $\bR$ directly on $Y$,
even if the ambient space $X$ is singular.
This leads to a natural proposal for a McKay correspondence 
for quotient singularities that do not have a complete, crepant
resolution. In particular there is still a correspondence 
between the group theoretical tensor product \gisf\ and the 
intersections of the elements of the K-theory group on
a smooth hypersurface $Y$ in the maximal crepant resolution 
of $\CC^{n+1}/\Gamma$ defined by the LSM.

\subsec{The geometric setup from a string point of view}
As we will focus from now on more concretely to D-branes
in type IIA compactification on Calabi--Yau manifolds, 
let us briefly summarize the geometric setup. The starting point is
the by now familiar quotient singularity $Z=\CC^{n+1}/\Gamma$, 
with $\Gamma\subset SU(n+1)$ a discrete group 
such that there is a (partial) resolution $\tx Z$ of 
$Z$ with trivial canonical bundle. 
The space $\tx Z$ may be relevant for string theory in two 
very different ways. Firstly, $\tx Z$ may appear as a local patch of
a \CY $n+1$-manifold, on which the string propagates, e.g.
a K3 manifold in the case $\CC^2/\Gamma$. Secondly, we may 
embed a Calabi--Yau $(n-1)$-fold $Y$ as a hypersurface 
(or intersections thereof) in the compact exceptional divisor $X\subset
\tx Z$. In this case the string 
propagates only on $Y$. 

The two cases are closely related: to construct the Calabi--Yau $Y$ 
we consider first the total space of the anti-canonical bundle 
$\cx K^*_X$. This is precisely the (partial) resolution of 
the non-compact \CY $Z$. The hypersurface $Y$ is then defined 
as the zero locus of a generic section of $\cx K^*_X$. E.g.,
for the quintic, $\tx Z$=$\cx O(-5)_{\IP^4}$ and $Y$ is the zero locus
of a quintic polynomial $p(x_i)$ in $X=\IP^4$. 

We may obtain elements of the K-theory group $K(Y)$
by restricting the large volume version of the
bases $\bR$ and $\bS$ to the hypersurface \rDD.
This map is not bijective for small $n$ but becomes better
with increasing $n$ \ref\rMDpc{M. Douglas, private communication.}. 
A surprising  consequence is that 
{\it the D-brane states  on $Y$ are
to a large extent determined by the the representation theory of 
a simple orbifold}. We will describe
in the next section\fff how detailed information, such 
as monodromy matrices and analytic continuation of periods,
may be extracted from the K-theory data defined in this way.
Note that the orbifold limit Vol$(X) \to 0$ 
implies also a small volume limit of the hypersurface $Y$.

\subsec{A McKay correspondence for singular resolutions}
The question inverse to the restriction of $K(X)$ to $K(Y)$ is:
{\it to what extent do the K-theory data on $Y$
describe a McKay correspondence} ? This becomes important in 
dimensions higher than three, as not all the singularities of the
form $Z=\CC^{n+1}/\Gamma$ with $\Gamma\subset SU(n+1)$ 
allow for a resolution that
keeps the canonical bundle trivial and a generalization of 
the McKay correspondence is not obvious. However, a Calabi--Yau 
hypersurface
embedded in a partial resolution of $Z$ may avoid the remaining
singularities. We will argue now that 
the bases $\{R_a\}$ and $\{S^a\}$ can be specified entirely on the 
hypersurface $Y$ embedded in the partial resolution defined by the 
linear sigma model. For $n>2$ this gives a concrete proposal for a McKay
correspondence, even if the maximal, crepant resolution of $Z$
is still singular: we may define the McKay correspondence as a 
relation between the group theory of $\CC^{n+1}/\Gamma$ and the 
compact K-theory group $K_c(H)$ generated by $\bS$ on a
smooth, compact hypersurface $H$ of minimal codimension in 
the maximal, crepant resolution $\tx Z$ of $Z$ as defined by 
the LSM.

Let us consider again the inner product \HRR. 
The Todd classes of $X$ and $Y$ are related by
\eqn\toddi{
\td(X)={1\over 2}\, c_1(X)\, \td(Y)+\rmx{even},}
where the term (even) denotes that it contains only
$(2k,2k)$-forms. 
From $\int_Y(.)=\int_X c_1(X) (.)$, we have
\eqn\isrels{\eqalign{
\la A,B \ra _X &= \fc{1}{2}\, \la A,B \ra _Y +X(A,B),\cr
\la B,A \ra _X &= \fc{1}{2}\, \la A,B \ra _Y \, (-)^{n-1}+
X(A,B)\, (-)^n,\cr}}
where $n=\rmx{dim}_{\CC}(X)$ and the form of $X(A,B)$ is irrelevant
up to the fact that it transforms with a sign $(-)^n$ under the exchange of
$A$ and $B$. Note also that $\td(Y)$ contains only $(2k,2k)$-forms,
which implies that $\la A,B \ra _Y$ transforms with a sign $(-)^{n-1}$ under
the same exchange.
It follows that 
\eqn\isrelii{
\la A,B\ra_Y=\la A,B\ra_X+(-)^{n-1}\, \la A,B\ra_X.}
If $\la A,B\ra_X$ is upper triangular, as is the 
case for $A,B$ are elements of $\bR$ or $\bS$, 
it is completely determined in terms of 
$\la A,B\ra_Y$:
\eqn\isreliii{
\chi_{ab}=\la R_a,R_b\ra_X=\cases{
\la R_a|_Y,R_b|_Y \ra_Y,\ & $b>a$\cr 
\quad 1,\ & $b=a$\cr 
\quad0,\ & $b<a$}.}
The formulae \isrelii\ and \isreliii\ 
allow to go forth and back between the data on $Y$ and 
that on $X$. Note that the inner product 
$\la R_a,R_b \ra_Y$ is still completely determined
by the representation theory of $\Gamma$; it is not invertible on 
$Y$, however. The restrictions $V^a=S^a|_Y$ in terms of the data on $Y$ 
are
\eqn\vdefs{
V^{a}=\mchi^{ab}R^*_b|_Y,}
with $\chi^{ab}$ the inverse of the matrix defined in \isreliii.
The intersections of the $V^{a}$ on $Y$ are related to that on $X$
by \isrelii, $\la V^a ,V^b \ra _Y= \chi^{ab}+(-)^{n-1}\chi^{ba}$.
This completes the relation between the representation theory of 
$\Gamma$ and the K-theory group with compact support on $Y$.
We refer to appendix C for a simple example, where the LSM does not
provide a smooth resolution and thus the ambient space $X$ is singular.

\newsec{Application to Calabi--Yau three-folds}
In this section we apply the previous ideas to study stringy D-branes
on a Calabi--Yau 3-fold $Y$. The investigation of this subject was
initiated in \rDQ\ for the quintic hypersurface in $\IP^4$, 
falling back on previous work on boundary states in CFT \rRS\rNW. 
See also \rDG\rCYDbii\rSch\rDFR\rCYDbscft\rGJlsm\rCYDbsTG\ for subsequent studies
of this subject.

The main focus of this section is to demonstrate that the representation
theory of the embedding singularity contains a surprising amount
of information about the closed string sector on the compact
Calabi--Yau manifold embedded in it.
The main ingredient is the map between D-branes at small and large volume
described in the previous sections which, apart from being much
simpler then the approaches used so far, allows to reconstruct 
detailed data of the closed string sector, namely 
intersections, the large volume prepotential, 
monodromies and analytic continuation matrices,  from the 
simple group theoretical data of the open string sector.

\subsec{Fractional branes}
The natural, fundamental objects of the type II string at small volume
are the D-branes wrapped on the compact cohomology of $X$, 
the so-called fractional
branes. They correspond to the generators $S^a$ of the
basis of $K_c(X)$. In \rDD\ it was conjectured, that the restrictions of the
fractional brane states to a Calabi--Yau $Y$ embedded as a 
hypersurface in $X$, represent the rational B-type boundary 
states of a Gepner model that describes the small volume 
limit of $Y$. 
Let us give an explicit expression for 
the intersection form of the restriction $V^a=S^a|_Y$ of the 
fractional branes to a 3-fold hypersurface $Y$.
From \isrelii\ we obtain: 
\eqn\isV{\eqalign{
I_{LG}^{ab}&=\la V^a,V^b \ra_Y=\mchi^{ab}-\mchi^{ba}\cr
&=\Big(\prod_i(1-h^{w_i})-\prod_i(1-h^{w_i})^T\Big)^{ab}\ = \ 
\Big(\prod_i(1-g^{w_i})\Big)^{ab}.}}
Here $g$ is the extended shift matrix $g=h+\delta_{N,1}$.
The notation indicates that this intersection form is to be
identified with that of the LG states, according to the conjecture.
This quantity may be calculated if the theory has a representation
as a Gepner model \rDQ. The result is precisely the last expression
in \isV. This proves that the intersection matrices of the fractional branes
and the LG boundary states is the same and 
gives strong evidence for their identification.
In fact the result \isV\ does not depend on the existence
of a Gepner model and generalizes to all hypersurfaces
in weighted projective space.

\subsec{An open string derivation of the large volume prepotential}
The integral Chern classes of the fractional branes may be determined from 
\vdefs\ by simple linear algebra. The other
preferred, integral lattice that labels the charges of the D-brane states
is the symplectic lattice of BPS charges in the closed string theory
discussed in sect. 5\fff. The classes of the fractional branes 
on the 3-fold $Y$ and their integral, symplectic charges $\Qv$ are 
related by
\eqn\centralc{
Z(A)=-\int_Y\, e^{-J}\, \ch(A) \, \sqrt{\td(Y)}= \Qv\cdot \Piv,}
where $J=\sum_{\al=1}^{\hoo}\, t_\al\, K_\al$ is the K\"ahlerform on $Y$
and $A$ an element of the K(Y). The expression on the left hand
side\foot{This K-theoretic expression and the 
more conventional form involving the Mukai vector of a bundle are
related by a Grothendieck-Riemann-Roch argument, as explained in 
\rDR.} is derived from the coupling of the D-brane world volume
theory to the to the background fields \rDBcoups\
and the right hand side is the 
well-known central charge formula in the closed string theory \CAFP.
Equating the central charge of the D-brane states
as obtained from the open or closed string picture, respectively,
leads to the above relation \rDQ\rDR.

The symplectic basis is specified by the 'period vector' $\Piv$ which is
the section of a $SL(2\hoo+2,\ZZ)$ bundle. Its general form is 
determined by $\cx N=2$ space-time supersymmetry in terms of a 
holomorphic prepotential $\cx F(t_\al)$:
\eqn\periodv{
\Piv(t_\al)=\pmatrix{\Pi_6\cr \Pi_4^\al\cr\Pi_0\cr\Pi_2^\al}=
\pmatrix{2\cx F-t_\al\cx \p_\al \cx F \cr \cx \p_\al \cx F \cr 1 \cr t_\al}.}
It is easy to see that the large volume form of the prepotential $\cx F$
is determined by the open string formula \centralc\ 
and we need not to invoke mirror symmetry to determine it\foot{See
\rms\rTis\ and references therein for the derivation of $\cx F$  
from mirror symmetry and \rHos\ for a related discussion.}. 
The leading terms
of the central charge $Z$ for a 6-brane and a 4-brane that is 
wrapped on the divisor $E_\al$ dual to $K_\al$ are
\eqn\sixb{
Z_{6}=\int_Y \fc{J^3}{3!}+J\, \fc{c_2(Y)}{24},
\qquad Z_4=\int_Y \fc{-J^2\, K_\al}{2}+\fc{J}{2}\, i_*c_1(E_\al).}
Here $i$ denotes the embedding $i:E\hookrightarrow Y$ and 
we have used the Grothendieck-Riemann-Roch formula
to relate the K-theory class on $Y$ and the Mukai vector of the 
bundle on $E$. It follows that 
the large volume form of the prepotential is related to the topological
intersection data on $Y$ by 
\vskip 10pt
\vbox{
\eqn\prepot{
\cx F= -\fc{1}{3!}\, C_{\al\be\ga}t_\al t_\be t_\ga +
\fc{1}{2!}t_\al t_\be A_{\al\be} +
B_\al t_\al,}
\vskip-10pt
\ni with
\vskip-10pt
\eqn\asd{\eqalign{
C_{\al\be\ga}&=\int_Y K_\al K_\be K_\ga,\qquad 
B_\al =\fc{1}{24}\, \int_Y K_\al\, c_2(Y)\ \rmx{mod}\ \ZZ, \cr
A_{\al\be}&=\fc{1}{2}\int_Y K_\be\, i_*c_1(E_\al) \ \rmx{mod}\ \ZZ.
}}}
\vskip-16pt \ni
A simple check of these expressions is given by the requirement
that the period vector $\Piv$ transforms by a symplectic 
transformation under the shifts of the $B$-field, $t_\al \to t_\al+1$.
This implies $A_{\al\be}+\fc{1}{2}C_{\al\al\be}\in \ZZ,
\ 2B_\al+\fc{1}{6}C_{\al\al\al} \in \ZZ$ which is 
guaranteed by the relations $2B_\al = -\fc{1}{6}C_{\al\al\al}+\chi(E_\al)\ \rmx{mod}\ \ZZ$
and $A_{\al\be}=-\fc{1}{2} C_{\al\al\be}\ \rmx{mod}\ \ZZ$.

From eqs. \periodv\ and \centralc\ we may then obtain the relation 
between the Chern classes and the closed string charges $\Qv$ 
for any brane configuration.

\subsec{From open to closed strings: a representative example}

In the following we will use the information from the open
string sector, namely the fractional branes, to obtain 
detailed information on the moduli space of the closed string sector
on $Y$. As a basic example note that we can determine
the intersection data \asd\ on $Y$ from the tensor product 
formula \gisb\ by expressing the K\"ahler classes $K_\al$ 
in terms of Chern classes of 
the generators $R_a$ of the K-theory group.

It seems useful to explain the following arguments on the basis
of a concrete example, which we choose to be 
the degree 24 hypersurface embedded in a
partial resolution of a $\CC^5/\ZZ_{24}$ singularity. 
The \CY manifold $Y$ has $\hoo=3$ and has a sufficient degree of 
complexity to serve as a representative example in many respects.
Firstly the smooth hypersurface $Y$ may be embedded in a singular ambient
space and provides an example for a McKay correspondence in a 
singular resolution\foot{In this case, however, the additional
singularities {\it do} have a crepant resolution 
in the linear sigma model, so that
the smooth hypersurface of minimal codimension in $X$ is 
the ambient space $X$ itself. This will lead to a trivial
linear dependence of the basis $\bS$ in $K(Y)$, 
as further discussed below.}.
Also, while it would be hard to approach with
the closed string methods, namely analytic continuation used in 
\rDQ%
\rDG
\rDR
\rCYDbii
\rSch%
\rDFR
\ or
the toric description of refs.\rDGM\rDD, it is easy to deal with from the
point of the open string picture and thus demonstrates the 
effectiveness of this framework. 

\subsubsec{The group theoretical input}
The discrete group is defined by the weight vector $w=(1,1,2,8,12)$
and acts as $x_i\to \om^{w_i} x_i$ on the coordinates, with $\om^{24}=1$.
A partial resolution of this space is provided by the total space
of the bundle $\cx O(-24)_X$, $X={\WP_{1,1,2,8,12}^4}$. The Calabi--Yau
hypersurface $Y$ is defined as the zero locus of a generic section
of this bundle.

The starting point for the construction of the fractional branes 
are the group theoretic formula \gisb\gisf\ for
the inner product of the elements in $\bR$,
and its inverse that describes classes the fractional branes $\bS$ 
and their intersections:

\eqn\gthdataii{\eqalign{
\chi_{ab}^{\ZZ_{24}}&=
\pmatrix{\ss{1\, 2\, 4\, 6\, 9\, 12\, 16\, 20\, 26\, 32\, 40\, 48\, 59\, 70\, 84\, 
98\, 116\, 134\, 156\, 178\, 205\, 232\, 264\, 296}\cr
\ss{0\,1\, 2\, 4\, 6\, 9\, 12\, 16\, 20\, 26\, 32\, 40\, 48\, 59\, 70\, 84\, 
98\, 116\, 134\, 156\, 178\, 205\, 232\, 264}\cr
\ddots\hskip3cm \vdots\hskip3cm\cr
\ss{0\,\   0\,\   0\,\   0\,\   0\,\   0\,\   0\,\   0\,\   0\,\   0\,\   0\,\   0\,\   0\,\   0\,\   0\,\   0\,\   0\,\   0\,\   0\,\   0\,\   0\,\   0\,\   0\   1}},\cr\cr
\chi^{\ZZ_{24}\, ab}&= 
\pmatrix{
\ss{1\  -2\  0\  2\  -1\  0\  0\  0\  -1\  2\  0\  -2\  
0\  2\  0\  -2\  1\  0\  0\  0\  1\  -2\  0\  2}\cr
\ss{0\ 1\  -2\  0\  2\  -1\  0\  0\  0\  -1\  2\  0\  -2\  0\  2\  0\  -2\  1\  0\  0\  0\  1\  -2\  0}\cr
\ddots\hskip3cm \vdots\hskip3cm\cr
\ss{0\,\   0\,\   0\,\   0\,\   0\,\   0\,\   0\,\   0\,\   0\,\   0\,\   0\,\   0\,\   0\,\   0\,\   0\,\   0\,\   0\,\   0\,\   0\,\   0\,\   0\,\   0\,\   0\   1}}.}}

\ni
Note that the $k$-th row is just a shift of the first row by $k-1$ to the right,
with zeros appended at the left. The matrix $\chi^{ab}$ specifies the classes
of the fractional branes $S^a$ in terms of that of the $R_a$
by \sdef. 

We may choose the elements $R_a$ in the foundation $\bR$ 
of the helix $\cx H_R$ to lie at the origin, 
$\bR=\{\cx O(-23\, K),\dots,\cx O\}$. Here $K$ denotes 
the hypersurface class of $X$.

\subsubsec{The toric resolution}
To proceed we perform a toric resolution of the singularities of the 
weighted projective space. A complete resolution 
is described by the fan spanned by the following vertices 
\eqn\verts{\eqalign{
&\ss
(0,0,0,-1),
(0,0,-1,0),
(0,0,2,3),
(0,-1,0,0),
(0,1,4,6),
(-1,0,0,0),
(1,2,8,12),\cr
&\ss(0,0,0,1),(0,0,1,1),(0,0,1,2).}}
However only the divisors associated to the vertices in the first line 
do intersect the generic hypersurface \rBat. As we have argued, by the relation 
\vdefs, we may define the bases $\bR$ and $\bS$ 
entirely on $Y$ and their intersections in terms of 
$\la A,B\ra_Y=\int_Y \ch{A^*}\ \ch{B}\ \rmx{Td}(Y)$. Therefore we will work 
on the singular space $\hx X$ with the singularities corresponding to the
last three vertices in \verts\ unresolved. 

In the large volume phase, the gauged LSM has the
gauge symmetry $U(1)^{\hoo=3}$ and matter fields with charges
\vskip 0.3cm

{\vbox{{
\eqn\mori{
\vbox{\offinterlineskip\tabskip=0pt\halign{\strut
\hfil~$\ss{#}$~\hfil\vrule 
&\hfil ~$\ss{#}$~
&\hfil~$\ss{#}$~
&\hfil~$\ss{#}$~
&\hfil~$\ss{#}$~
&\hfil~$\ss{#}$~
&\hfil~$\ss{#}$~
&\hfil~$\ss{#}$~
&\hfil~$\ss{#}$~\vrule
&\hfil~$\ss{#}$~\cr
&p&x_5&x_4&z&x_3&w&x_2&x_1\cr
\noalign{\hrule}
l_1&-6& 3& 2& 1& 0& 0& 0& 0&\times4\cr
l_2& 0& 0& 0&-2& 1& 1& 0& 0&\times2\cr
l_3& 0& 0& 0& 0& 0&-2& 1& 1&\times1\cr
\noalign{\hrule}
l_K&-24& 12& 8& 0& 2& 0& 1& 1\cr}}
\ .}
}}}

\ni
The vectors $l_\al$ span the K\"ahler cone of $Y$ in the large volume phase.
We have also indicated the single 
$U(1)$ charge vector $l_K=4\, l_1+2\, l_2+l_3$ 
of the fields in the LG phase. 

It will be useful to have a little understanding of the geometry 
of the manifold $Y$. The fields $x_i,\ i=1,\dots,5$ are the
original coordinates on the weighted projective space $\WP_{1,1,2,8,12}^4$.
The fields $z$ and $w$ are the extra fields introduced in the resolution and
associated to the exceptional divisors. The manifold $Y$ is an
elliptic fibration over the Hirzebruch surface $\bx F_2$. The elliptic
fiber $E$ is parametrized by the coordinates $(x_4,x_5,z)$, while
the Hirzebruch surface is the divisor 
$\bx F_2:\, z=0$, parametrized by $(x_1,x_2,x_3,w)$. 
The base $\bx F_2$ is itself
a $\IP^1$ fibration over $\IP^1$, where the former, denoted by $F$,
is parametrized by $(x_3,w)$ and the latter is the divisor 
$B:w=0$ in $\bx F_2$ with coordinates $(x_1,x_2)$.

\subsubsec{The basis $\bR$ of geometric bundles}
We have already specified the elements of $\bR$ in the LG phase.
From the $U(1)^3$ charges of the fields $X_i$ and the relation of the 
charge vectors $l_\al$ to $l_K$, we may reconstruct the classes of 
line bundles $\hx R_a$ on the partial resolution $\hx X$ as explained in 
sect. 5.3\fff:
\eqn\bundm{\eqalign{
\cx E_{\hx X}=\{
&\ss\cx O(-5,-1,-1),\, \cx O(-5,-1,0),\, 
\cx O(-5,0,-1),\, \cx O(-5,0,0),\, \cx O(-4,-1,-1),\, \dots\cr
&{\ss\dots,\, \cx O(-1,0
,-1),\, \cx O(-1,0,0),\, \cx O(0,-1,-1),\, 
\cx O(0,-1,0),\, \cx O(0,0,-1),\, \cx O(0,0,0)}\}\cr
&\qquad \matrix{\ss LG\cr\to\cr{\phantom 1}} \ 
\{{\ss\cx O(-23),\cx O(-22),\dots,\cx O}\}.
}}
The bundles $\hx R_a$ in $\cx E_{\hx X}$ 
descend to the ground states $R_a$ in the LG phase, as indicated.
Here we use the standard notation $[\cx O(a)]=a\, K$ and 
$[\cx O(a,b,c)]=a\, K_1+b\, K_2+c\, K_3$, where $K_\al$ are the
$(1,1)$ forms on $\hx X$ related to the generators $l_\al$.

We could perform a further resolution of $\hx X$ to $\tx X$ 
to blow up the remaining singularities in the ambient space.
We may reconstruct the basis $\{\tx R_a\}$ on $\tx X$
in the same way as above from the charges of the fields w.r.t. to  
the $U(1)^6$ gauge symmetry in this phase. This 
realizes our claim that we can go from the 
data on the resolutions to that on the singular space and {\it vice versa}.
However there is no purpose for us to do so and we will continue to work with
the partial resolution $\hx X$ that resolves only the singularities
on the hypersurface.

\subsubsec{The fractional branes $V^a$}
We will now describe the structure of the fractional branes $V^a$ 
on $Y$ and find complete agreement with the arguments in sect. 2.
The orbit of 24 fractional branes $V^a$ comes in two sets of
12, which describe the branes and anti-branes on $Y$. The 
reason for the simple linear dependence of the charges 
$V^a$ with $a>12$ on those with $a<12$ is the fact that $Y$
misses singularities of the ambient space $X$ which may be resolved.
In the resolution, there are new K-theory classes on $\tx X$ which 
lie in the kernel of the map $K(\tx X)\to K(Y)$. Note that the same
will not happen if $Y$ misses singularities that do {\it not} have 
a crepant resolution.

Let us first consider the 12 branes.
With $i:\, \bx F_2 \to Y$ denoting the inclusion, the
12 branes build on $V^1=\OO{(-1,1,1)}$ 
correspond to the sheaves $V^a=\OO{(-1,1,1)}\otimes \tx V^a$ with
\eqn\fbxtf{
\vbox{\offinterlineskip\tabskip=0pt\halign{\strut
$#=$~\hfil&$#$~\hfil\cr
\tx V^{1}&+\OO{(0,0,0)},\cr
\tx V^{2}&-\OO{(0,0,1)},\cr
\tx V^{3}&-\OO{(0,1,-2)},\cr
\tx V^{4}&+\OO{(0,1,-1)},\cr
}}
\ \ 
\vbox{\offinterlineskip\tabskip=0pt\halign{\strut
$#=$~\hfil&$#$~\hfil\cr
\tx V^{5}&-i_*\, \OO{(0,-2,0)},\cr
\tx V^{6}&+i_*\, \OO{(0,-2,1)},\cr
\tx V^{7}&+i_*\, \OO{(0,-1,-2)},\cr
\tx V^{8}&-i_*\, \OO{(0,-1,-1)},\cr
}}
\ \ 
\vbox{\offinterlineskip\tabskip=0pt\halign{\strut
$#=$~\hfil&$#$~\hfil\cr
\tx V^{9}&-\OO{(1,-2,0)},\cr
\tx V^{10}&+\OO{(1,-2,1)},\cr
\tx V^{11}&+\OO{(1,-1,-2)},\cr
\tx V^{12}&-\OO{(1,-1,-1)}.\cr
}}
}
\ni
Note that the sign in the definitions \fbxtf\ counts precisely the
fermion number of a map $V^{1}\to V^a$. 

To interpret the sheaves $V^a$ in the spirit of the previous sections,
note first, that the fermions $\psi^1,\dots,\psi^5$ live on
the whole space $Y\subset X=\WP^{4}_{1,1,2,8,12}$. As we discussed already, 
this a consequence of the fact that the associated super-fields 
parametrize the exceptional divisor $X$ of the first 
resolution of the orbifold $\CC/\ZZ_{24}$. On the other hand, 
the fermions $\zeta^z$ and $\zeta^w$ are related to 
resolutions of the singularities on $Y\subset X$. They introduce new 
boundary conditions that  correspond to branes wrapped on the
divisors $z=0$ and $w=0$, respectively. 

The first non-trivial bundle $V^{2}$ is obtained in the
LG phase by multiplication
of $\cx O$ with $\psi^{1,2}$. The corresponding bundle is described by 
the restriction of the sequence \fermo\ for the tangent bundle on $X$ 
to $B$. We may thus identify $V^{2}$ with the pull back of the
tangent bundle $\Omega_B^*(-K_3)$ to $Y$. Note that $\{V^{1},V^{2}\}$
is the pull back of the helix $\cx H_{\IP^1}=
\{\cx O,\Omega^*_{\IP^1}(-1)\}$ on $\IP^1$ to $Y$\foot{As 
the dimension of $B$ is one, we obtain only line bundles. See 
appendix B\fff for a completely analogue example in a 
less degenerate case.}.

The bundle $V^{3}$ is obtained in the LG phase 
by multiplication of $\cx O$ by $\psi^{3}$. In the geometric phase,
this field gets part of the tangent bundle of the sphere $F$, 
with sections $\psi^3,\zeta^w$. As the field $w$ is also
a section of $\cx O(-2\, K_3)$, the resulting bundle is the pull back of 
$\Omega_F^*(-K_2-2K_3)$ to $Y$, in agreement with \fbxtf.
Similar the bundle $V^{4}$ corresponds to 
multiplication of $\cx O$ by $\psi^{3}\psi^1$. From 
the above, we identify $V^{4}$ as the pull back of 
$\Omega^*_B\otimes \Omega^*_F(-K_2-3K_3)$.

The next set of four elements is more interesting as they correspond to 
sheaves on $\bx F_2$ extended by zero. Note that this is precisely
the situation where the map from $V^1\to V^5$ is gauge equivalent to
zero and moreover there is no map $V^{1}\to V^{a}$
in the LG phase for $a=6,7,8$. However, in the geometric phase, there is
the extra field $\zeta^z$ which generates non-trivial sections of
these sheaves. 
Multiplication by $\zeta^z$ sets to zero the
bosonic component $z$ and thus the new brane sits on the divisor
$\bx F_2:\, z=0$. This agrees perfectly with the result in \fbxtf !
We thus identify $V^{5}$ with $i_*\cx O_{\bx F_2}{\ss (0,-2,0)}$; the next
three sheaves are constructed precisely as in the first set of four
above and correspond to the pull backs of the bundles 
$\Omega_B^*(-2K_2-K_3)$, $\Omega_F^*(-3K_2-2K_3)$ and 
$\Omega_B^*\otimes\Omega_F^*(-3(K_2+K_3))$ to $\bx F_2$, further
extended by zero on $Y$.

As for the next set of four, there is again a fermion, namely $\psi^4$,
that maps $V^{1}\to V^{9}$ in the LG phase and yields thus
again a bundle on $Y$. Note that the bundle $V^a$ is defined as the
restriction of $S^a$ to $Y$ and thus $V^9$ is determined by the 
section associated with $\psi^4$ on $X$, not on $Y$.

Finally the set of 9 anti-branes is constructed in the very same way 
as the set of the 9  branes with the difference that there is one extra factor
$\psi^5$ in the definition of their sections, which gives
the minus sign for each $V^a$. 
We note also that the repetition pattern in groups of four 
is related to the shift 
of the origin of the dual foundation $\{R_a\}$ by four units. 
In particular we will show below that this 
shift corresponds  to the monodromy $t_1\to t_1+1$ which preserves 
the large volume limit in the Calabi--Yau  moduli.

\subsubsec{The BPS charge lattice}
It is straightforward to determine the topological data \asd\ of $Y$.
They are summarized by the following prepotential:
\eqn\prepot{
\cx F=-t_3 t_2 t_1-\fc{4}{3} t_1^3-t_3 t_1^2-t_2^2 t_1-2 t_1^2 t_2
+\fc{23}{6} t_1+2 t_2+t_3,}
up to terms constant and exponential in the $t_\al$. 
From \centralc\ we obtain the following $N\times 2\hoo+2$ 
matrix $Q_S$ which 
describes the symplectic charges $\Qv$ for the fractional branes 
$V^a=S^a|_Y$:
\vskip 0.4cm
{\vbox{{
\eqn\qv{\eqalign{
&\vbox{\offinterlineskip\tabskip=0pt\halign{\strut
\hfil~$\ss{#}$~\hfil\vrule&
 \hfil~$\ss{#}$
&\hfil~$\ss{#}$
&\hfil~$\ss{#}$
&\hfil~$\ss{#}$
&\hfil~$\ss{#}$
&\hfil~$\ss{#}$
&\hfil~$\ss{#}$
&\hfil~$\ss{#}$
\cr
&Q_6&Q_4^1&Q_4^2&Q_4^3&Q_0&Q_2^1&Q_2^2&Q_2^3\cr
\noalign{\hrule}
V^1&-1&1&-1&-1&-2&0&1&0\cr
V^2&1&-1&1&2&0&-1&-2&0\cr
V^3&1&-1&2&-1&2&0&-1&-1\cr
V^4&-1&1&-2&0&-1&0&2&1\cr
V^5&0&1&-2&0&0&0&1&0\cr
V^6&0&-1&2&0&0&0&-2&0\cr
V^7&0&-1&2&0&0&0&-1&-1\cr
V^8&0&1&-2&0&-1&0&2&1\cr
V^9&1&0&-1&1&2&0&0&0\cr
V^{10}&-1&0&1&-2&0&1&0&0\cr
V^{11}&-1&0&0&1&-2&0&0&0\cr
V^{12}&1&0&0&0&0&0&0&0\cr
}}\ \ ,\cr\cr
\ss &\qquad \ \ \vec{Q}(V^{12+a})=-\vec{Q}(V^{a}).}}}}}

\subsubsec{Monodromies}
%
From the symplectic charges of the fractional branes we may
derive further information about the monodromies of the
K\"ahler moduli space $\cx M_K$  of $Y$. Let us first consider the
monodromy around the divisor $C_0$ in $\cx M_K$ which corresponds 
to the LG point 
at small volume\foot{To be precise, $\cx M_K$ denotes enlarged 
K\"ahler moduli space which has been resolved such that the 
components of the discriminant intersect with normal crossings.}.
At this point, the moduli space has a $\ZZ_{N}$ monodromy that permutes 
cyclically the line bundles of a given foundation, 
$R_a\to R_{a-1}$ for $a>1$ and  $R_1\to R_N$. From 
the orthogonality relation between the $R_a$ and the $S^a$ it follows,
that the effect on the $S^a$ is the same. This transformation may
be realized as a left multiplication of $Q_S$ by the matrix $g^T$.
The fact that this transformation corresponds to a monodromy in the
moduli space implies the existence of a $(2\hoo+2,2\hoo+2)$ matrix 
$A$ that fulfills $g^T\cdot Q_S=Q_S\cdot A$.
It is straightforward to determine the matrix $A$ from \qv:
\vskip12pt
\vbox{\ninepoint{
\eqn\monoii{
A=\pmatrix{-1&0&0&1&-2&0&0&0\cr0&1&0&0&0&2&1&0\cr1&-1&3&0&1&1&-2&-1\cr
0&0&1&-1&0&1&0&0\cr1&0&0&0&1&0&0&0\cr-1&1&-2&0&-1&1&2&1
\cr1&-2&4&0&1&0&-3&-1\cr1&0&0&0&1&0&0&-1}.}}}
\vskip-12pt
\ni
We may also consider the action on $Q_S$ under a shift of the origin of 
the foundation $\bR$ by one to the left. From the previous 
discussion we know that a shift by $N$ represents multiplication of
the basis $\bR$ by $c_1(M)$. However we do expect that even a single
shift represents a monodromy around a component $C_\infty$
of the discriminant locus, 
as the origin of the foundation 
$\bR$ should not be distinguished. Indeed we find that the
action $\{R_{n_0+1},\dots,R_{n_0+N}\} \to \{R_{n_0},\dots,R_{n_0+N-1}\}$
has a representation by right multiplication of $Q_S$ with a matrix
$T_{\infty}^{-1}=(AT)^{-1}$, where $T$ is the matrix
\vskip8pt
\vbox{\ninepoint{
\eqn\conifm{ 
T=\pmatrix{
1&0&0&0&0&0&0&0\cr0&1&0&0&0&0&0&0\cr0&0&1&0&0&0&0&0\cr0
&0&0&1&0&0&0&0\cr-1&0&0&0&1&0&0&0\cr0&0&0&0&0&1&0&0\cr0&
0&0&0&0&0&1&0\cr0&0&0&0&0&0&0&1}.}
}}
\vskip-8pt
\ni
We recognize $T$ as the monodromy of the conifold point, where the 
6-cycle corresponding to the whole space $Y$ shrinks to zero size.
We have checked the above relation between the unit shift of
the origin of $\bR$ and the monodromy $AT$, with $T$ the monodromy
of the zero size 6-brane in many other examples which suggests that
there should be a general reason for this universal behavior.

Let us also consider the action of $T_\infty$ on the K\"ahler coordinates
$t_\al$. Whereas the monodromy $AT$ does not commute with the distinguished large 
volume limit, its fourth power does:
\vskip12pt
\vbox{\ninepoint{
\eqn\atmonoii{
\pmatrix{t_1\cr t_2\cr t_3}\ {(AT)\atop \to}
\pmatrix{t_3 t_2+t_2^2-t_1-2t_2-t_3+\fc{7}{6}\cr
-2t_3t_2-2t_2^2+3t_2+t_3-\fc{4}{3}\cr
-1+t_3}
{(AT)\atop \to}
\pmatrix{-1+t_1+t_2\cr 1-t_2\cr t_3}{(AT)^2\atop \to} 
\pmatrix{-1+t_1\cr t_2\cr t_3}.}
}}
\vskip-12pt
\ni
In particular we see that a shift by $N=24$ corresponds to a
shift of the $B$-field by $c_1(M)=6\, K_1$, in full agreement
with the results obtained in \rHIV\ from local mirror
symmetry. We note also that the monodromy action $(AT)^2$
induces a $\ZZ_2$ symmetry of the instanton expansion of $Y$, as it 
preserves a large volume limit but acts non-trivially on the
$t_\al$.

\subsubsec{Analytic continuation from the Gepner point to large volume}
A major obstacle to reconstruct the global structure of the moduli
space $\cx M_K$ is to determine the precise linear combination of 
local solutions to differential equations satisfied by the 
periods that define a symplectic section $\Piv$. One way to
find this relation, used first in \rcand\ for the quintic with $\hoo=1$,
is the analytic continuation of period integrals. It is very hard
to generalize this technique to the case of larger $\hoo$. The open
string approach gives an effective way to determine 
the analytic continuation from the LG point to large volume
by simple linear algebra.

At the LG point, 
the $N$ D-brane states $V^a$ come in an orbit of the $\ZZ_N$ symmetry
with degenerate masses $\sim |Z(V^a)|$. One may use
$2h^{1,1}+2$ of the central charges $\om_a=Z(V^a)$ as a basis for the 
period vector $\Piv$ at the LG point and we may choose $\Piv_G=(\om_0,\dots,
\om_{2\hoo+2})$. Note that this is not yet a symplectic
section of $SL(2\hoo+2,\ZZ)$, but related to it by a linear transformation.
The remaining periods $\om_a$ may be expressed in terms 
of those in $\Piv$. One way to determine these relations is 
to study explicit expressions of period integrals as in \rcand. 
However there is a simpler way, given the intersection matrix 
$I^{ab}_{LG}=\chi^{ab}-\chi^{ba}$: the relations correspond simply 
to its zero vectors. 
There are $N-(2\hoo+2)$ of them and they take the following form
in the present example:
\skip 0.2cm

{\vbox{{
\eqn\erels{
\vbox{\offinterlineskip\tabskip=0pt\halign{\strut
~${#}$~\hfil\quad 
&~${#},$~\hfil\quad 
&~${#}$~\hfil 
\cr
\om_9 = -\om_1+\om_5,
& \om_{10} = -\om_2+\om_6
& \om_{11} = -\om_3+\om_7,\cr 
\om_{12} = -\om_4+\om_8,
& \om_{13} = -\om_1
& \om_{14} = -\om_2,\cr 
\om_{15} = -\om_3,& 
\om_{16} = -\om_4& 
\om_{17} = -\om_5,\cr 
\om_{18} = -\om_6,& \om_{19} = -\om_7& 
\om_{20} = -\om_8,\cr \om_{21} = \om_1-\om_5,& \om_{22} = \om_2-\om_6
& \om_{23} = \om_3-\om_7,\cr 
\om_{24} = \om_4-\om_8.\cr}}
}}}}

\ni
We may now determine the analytic continuation matrix $M$ between
the central charges $Z(V^a)$ and the large volume periods $\Piv$
by the formula 
\eqn\eac{
Q_S\cdot M=(1-g)\cdot \pmatrix{\rmx{\bx 1}_{2\hoo+2}\cr R},
}
where $R$ is the matrix of relations in \erels. The factor $(1-g)$
takes into account the fact that the LG states correspond to 
semi-periods, rather than periods of the LG model, see ref. \rSemip\
and sect.5.2 of ref.\rHIV.
In this way we find the searched for analytic continuation matrix
\eqn\eacmat{
\vbox{\ninepoint{
$$
M=\pmatrix{
-1&1&0&0&0&0&0&0\cr2&2&2&2&0&0&-1&-1\cr2&1&1&1
&-1&0&0&0\cr1&0&1&0&0&0&0&0\cr1&0&0&0&0&0&0&0\cr-1&0&0&0
&1&0&0&0\cr1&0&0&0&-1&0&1&0\cr1&0&0&0&-1&1&-1&1}.
$$
}}
}

\ni
{\it Note added:} On the date of publication we received
the preprints \rigsm\ which discuss similar issues and
have a certain overlap with the results in sect.7 and sect.9.\foot{The
above results imply conjecture $2a$ in the first paper 
and are in conflict with the conjecture $1$, 
as the generator $T_\infty$ defined in  
sect. 7.2. is not a large volume monodromy.}

\ni
{\bf Acknowledgments}: I am very grateful to Mike Douglas and
Wolfgang Lerche
for important remarks. I would also like to thank 
Emanuel Diaconescu,
Sheldon Katz,
Andy Lutken,
Yaron Oz,
Christian R\"omelsberger,
Christof Schmidhuber and 
Johannes Walcher
for valuable discussions and the Rutgers
physics department for hospitality during completion of this work.

\appendix{A}{A simple example}
Let us exemplify the above construction on the basis of
the simple example $\CC^5/\ZZ_5$ with 
$\cx O(-5)_{\IP^4}$ as the partial and complete resolution. The
$\ZZ_5 \subset U(1) \subset \CC^*$ acts on the homogeneous coordinates 
of $\IP^4$ as $x_i\to \om x_i$, with $\om^5=1$. 
The irreps $\rho_a$ of $\ZZ_5$ transform 
as $\om^a,\, a=0,\dots,4$ There are 5 linear monomials
generating the maps from $\rho_a$ to $\rho_{a+1}$,
15 quadratic monomials generating the maps $\rho_a$ to $\rho_{a+1}$,
and so on. The bilinear form $\chi^H_{ab}$ and its inverse, obtained 
from \gisb,\gisf, are
\eqn\pfis{\mchi^{\ZZ_5}_{ab}=
\pmatrix{1&5&15&35&70\cr0&1&5&15&35\cr0&0&1&5&15\cr0&0&0&1&5\cr0&0&0&0&1},
\ 
\mchi^{\ZZ_5\, ab}=\pmatrix{
1&-5&10&-10&5\cr
0&1&-5&10&-10\cr
0&0&1&-5&10\cr
0&0&0&1&-5\cr0&0&0&0&1}.}
With the choice $\{R_a\}=\{\cx O(-4),\cx O(-3),\dots,\cx O\}$
we obtain from \sdef\ the following 
Chern characters of the duals $S^a$:
\eqn\sfq{
\vbox{\offinterlineskip\tabskip=0pt\halign{\strut
\hfil~$#$~\vrule\quad&
\hfil~$#$\ &
\hfil~$#$\ &
\hfil~$#$\ &
\hfil~$#$\ &
\hfil~$#$\cr
&r
&\rmx{ch}_1\, [K^1]
&\rmx{ch}_2\, [K^2]
&\rmx{ch}_3\, [K^3]
&\rmx{ch}_4\, [K^4]\cr
\noalign{\hrule}
S^1  &1&-1\ \ \, & \fc{1}{2}\ \ \, &-\fc{1}{6}\ \ \, & \fc{1}{24}\ \ \,\cr  
-S^2 &4&-3\ \ \, & \fc{1}{2}\ \ \, & \fc{1}{2}\ \ \, &-\fc{11}{24}\ \ \,\cr  
S^3  &6&-3\ \ \, &-\fc{1}{2}\ \ \, & \fc{1}{2}\ \ \, & \fc{11}{24}\ \ \,\cr  
-S^4 &4&-1\ \ \, &-\fc{1}{2}\ \ \, &-\fc{1}{6}\ \ \, &-\fc{1}{24}\ \ \,\cr  
S^5&1&0\ \ \, & 0\ \ \, & 0\ \ \, & 0\ \ \,\cr
}}}
Upon restriction to the quintic Calabi--Yau manifold $Y$ described by the 
quintic $p_Y=\sum_{i=1}^5\, x_i^5 =0$,  
these expressions agree with the result in \rDQ, where 
the fractional branes have been identified using the cumbersome
analytic continuation of period matrices performed in 
ref.\rcand. The above Chern characters agree with that of the bundles\foot{The
following agreement was independently noted by A. Lutken.}
\eqn\sfqe{
S^{a}=(-1)^{a-1}\Lambda^{a-1} \Omega^*(-a),\ a=1,\dots,5.}

\appendix{B}{Fractional branes on the manifold $\WP^{4}_{1,1,1,6,9}[18]$}
Here we give another instructive example for the structure 
of the fractional branes. It is the case of the degree 18 hypersurface 
$\WP^{4}_{1,1,1,6,9}[18]$ with $\hoo=1$ considered in \rDR, based on 
the earlier study of \rcandii. We want to 
describe the fractional branes in terms of tangent bundles 
on submanifolds of $Y$. The manifold
$Y$ is a simple elliptic fibration over a $\IP^2$, which we denote by $E$. 
Due to the fibration structure, the fermionic zero modes associated
with $E$ describe literally the tangent bundle on  $\IP^2$. 
Let $i:E\to Y$ denote again the inclusion,
$\pi:X\to E$ the fibration 
and $\cx L(a,b)$ the twist of the bundle $\cx L$ by the line bundle
with Chern classes $a\, K_1+b\, K_2$. The 18 fractional branes $V^a$
come in a set of 9 branes and their anti-branes. To be precise, this
is only true for the restrictions $V^a=S^a|_Y$, but not for the
branes on the total space. We take, as before, a foundation $\bR$
left to the origin, that is $R_{18}=\cx O$. Then the first nine
fractional branes are the sheaves $V^a=\cx O(-1,2)\otimes \tx V^a$ with
\eqn\fbxet{
\vbox{\offinterlineskip\tabskip=0pt\halign{\strut
$#=$~\hfil&$#$~\hfil\cr
\tx V^{1}&+\cx O(0,0),\cr
\tx V^{2}&-\pi^*\Omega^*_{\IP^2}(0,-1),\cr
\tx V^{3}&+\cx O(0,1),\cr}}
\ \
\vbox{\offinterlineskip\tabskip=0pt\halign{\strut
$#=$~\hfil&$#$~\hfil\cr
\tx V^{4}&-i_*\, \cx O(0,-3),\cr
\tx V^{5}&+i_*\, \Omega^*_{\IP^2}(0,-4),\cr
\tx V^{6}&-i_*\, \cx O(0,-2),\cr}}
\ \
\vbox{\offinterlineskip\tabskip=0pt\halign{\strut
$#=$~\hfil&$#$~\hfil\cr
\tx V^{7}& -\cx O(1,-3),\cr
\tx V^{8}&+\pi^*\Omega^*_{\IP^2}(1,-4),\cr
\tx V^{9}&-\cx O(1,-2).\cr}}
}
We see that the fractional branes come in blocks of three 
which originate from the foundation 
$\cx E_S=\{\cx O,\Omega^*(-1),\wedge^2\Omega^*(-2)=\cx O(1)\}$ 
of the helix on $\IP^2$. 
The first block $\{V^{1},V^{2},V^{3}\}$ is 
the pull back of $\cx E_S$ to $Y$, 
the second one, $\{V^{4},V^{5},V^{6}\}$, is the extension of 
$\cx E_S \otimes \cx O(0,-3)$ by zero and the third is again a pull back
twisted by $\cx O(1,-3)$. The twist bundles are easily recognized 
as the charges of the matter field $z$  which describes the 
embedding of $E$ as a divisor in $Y$, $E:z=0$. They describe the
normal bundle of $E$ in $Y$.

The pattern of the repetition of the exceptional collection  $\cx E_S$ is related to the
shift of the $B$-field by $K_1$. In fact the mutation 
$\rmut^{N-1}R_1$ that shifts the origin by one to the right is
the monodromy 
\vskip8pt
\vbox{\ninepoint{
\eqn\conifmxet{ 
T^{-1}_\infty=\pmatrix{
1&-1&2&3&2&1\cr
0&1&0&1&-3&-1\cr
0&0&1&0&-1&0\cr
0&0&0&1&0&0\cr
0&-1&3&0&1&1\cr
0&3&-9&1&0&-2},
}\vskip-8pt
}}
\ni
which may be again written as $T_\infty=AT$, where $T$ is the 
monodromy of a massless D6-brane as in \conifm, and $A$ the monodromy
at the Gepner point. In particular
the monodromy $T_\infty^3$ for 
the shift of the origin of $\bR$ 
by three units preserves the large volume limit and
coincides with the shift of the B-field, $t_1\to t_1+1$.

\appendix{C}{A quotient singularity without crepant resolution}
A simple example where the LSM resolution does not provide a
complete crepant resolution is the quotient singularity 
$Z=\CC^5/\ZZ_6$ specified by the weight vector $w=(1,1,1,1,2)$.
In this case the index \HRR\ is not well-defined and we work
instead directly on the degree six hypersurface $Y$ embedded in the
exceptional divisor $X=\IP^4_{1,1,1,1,2}$. The index on the
hypersurface is given by eqs.\gisb,\gisf\ and  \isrelii\ and
takes the following form on the bases $\bR$ and $\bS$:

\eqn\appCi{\eqalign{
I_{LG\, ab}&=
\pmatrix{
0& 4& 11& 24& 46& 80\cr -4& 0& 4& 11& 24& 46\cr -11& -4& 0& 4& 11& 24\cr -24
& -11& -4& 0& 4& 11\cr -46& -24& -11& -4& 0& 4\cr -80& -46& -24& -11& -4& 0},
\cr\cr%
I^{ab}_{LG}&=
\pmatrix{
0& -4& 5& 0& -5& 4\cr 4& 0& -4& 5& 0& -5\cr -5& 4& 0& -4& 5& 0\cr 0& -5& 4& 0
& -4& 5\cr 5& 0& -5& 4& 0& -4\cr -4& 5& 0& -5& 4& 0}.}}
The generators $R_a$ in the large volume limit of $Y$  coincides with the 
definition in the LG phase, $\bR=\{\cx O(-5),\dots,\cx O\}$. Note that
the definition of the basis $S^a$, as described
in sects. 5.4 and 7.5\fff, is in terms of sequences on $Y$, as the
original sequences on the singular $X$ will not be exact. 
\listrefs
\end